\newcommand{\be}{\begin{equation}}
\newcommand{\bea}{\begin{eqnarray}}
\newcommand{\ee}{\end{equation}}
\newcommand{\eea}{\end{eqnarray}}

\def\x{\mbox{{\bf x}}}

\def\Y{\mbox{$\bf{Y}$}}
\def\X{\mbox{$\bf{X}$}}

\def\p{\mbox{$\bf{p}$}}

\RequirePackage{fix-cm}
\documentclass[twocolumn]{svjour3}          % twocolumn
\usepackage{graphicx}
\usepackage{epstopdf}

\smartqed  % flush right qed marks, e.g. at end of proof
\usepackage{fixltx2e}
\usepackage{amsmath}
\begin{document}

\title{Nonlinear Statistical Data Assimilation for HVC\textsubscript{RA} Neurons in the Avian Song System%\thanks{Grants or other notes
}
%\title{Nonlinear Statistical State and Parameter Estimation of HVC\textsubscript{RA} Neurons in the Avian Song System}


\author{Nirag Kadakia \and Eve Armstrong \and Daniel Breen \and Uriel Morone \and Arij Daou \and Daniel Margoliash \and Henry D.I. Abarbanel}


\institute{N. Kadakia\at
	   Department of Physics, University of California, 9500 Gilman Drive, La Jolla, CA 92903-0402 \\
              \email{nkadakia@physics.ucsd.edu}           \\
           \and
           E. Armstrong \at
	BioCircuits Institute, University of California, 9500 Gilman Drive, La Jolla, CA 92903-0328 \\
           \and
	D. Breen \and U. Morone \at
	   Department of Physics, University of California, 9500 Gilman Drive, La Jolla, CA 92903-0402 \\
           \and
           A. Daou \and D. Margoliash \at
	Department of Organismal Biology and Anatomy, University of Chicago, 1027 E. 57th Street, Chicago, IL 63607 \\
           \and
           H. D. I. Abarbanel \at
	Marine Physical Laboratory (Scripps Institution of Oceanography), Department of Physics, University of California, 9500 Gilman Drive, La Jolla, CA 92903 \\
}

\date{Received: date / Accepted: date}

\maketitle

%%%%%%%%%%% 			Abstract 			%%%%%%%%%%

\begin{abstract}
With the goal of building a model of the HVC nucleus in the avian song system, we discuss in detail a model of HVC\textsubscript{RA} projection neurons comprised of a somatic compartment with fast Na$^+$ and K$^+$ currents and a dendritic compartment with slower Ca$^{2+}$ dynamics. We show this model qualitatively exhibits many observed electrophysiological behaviors. We then show in numerical procedures how one can design and analyze feasible laboratory experiments that allow the estimation of all of the many parameters and unmeasured dynamical variables, given observations of the somatic voltage $V_s(t)$ alone. A key to this procedure is to initially estimate the slow dynamics associated with Ca, blocking the fast Na and K variations, and then with the Ca parameters fixed, estimate the fast Na and K dynamics. This separation of time scales provides a numerically robust method for completing the full neuron model, and the efficacy of the method is tested by prediction when observations are complete. The simulation provides a framework for the slice preparation experiments and illustrates the use of data assimilation methods for the design of those experiments.
\end{abstract}

%%%%%%%%%%% 			Introduction 			%%%%%%%%%%

\section{Introduction}
\label{intro}
Biological neural circuits comprise collections of cells with distinct anatomical features interwoven via synaptic connections~\cite{Scotland,Johnston_Wu}. To build biophysically realistic models of functional networks, one requires knowledge of individual ion channel responses and intracellular processes contributing to postsynaptic behavior. To then reliably place these nodes into a functional network, one must then establish the connectivity among different classes of neurons. 

This paper and its companions are directed toward the construction of the network of the HVC nucleus in the song production pathway of zebra finch songbirds. Male zebra finches express a short, stereotyped song motif throughout their adult lifetime, following a several month sensorimotor period of learning and perfecting the song as juveniles~\cite{Hahnloser_2002}. The firing patterns of the neurons in the song production circuit are invariant through repeated renditions of this song~\cite{Konishi}. Within this microcircuit, neurons which project from HVC to the robust nucleus of the arcopallium (RA), HVC\textsubscript{RA} neurons, exhibit short bursts at most once during a motif~\cite{Hahnloser_2002}.

Here we investigate the following: 
\begin{itemize}
  \item a model for HVC\textsubscript{RA} neurons that reproduces observed behavior, and \\
  \item an experimentally realizable protocol for estimating and predicting the parameters and unmeasurable states of the model, utilizing improved methods of statistical data assimilation~\cite{Abarbanel_book}.  
\end{itemize}  

The data assimilation procedure is an estimation and validation method for unknown parameters and unobserved state variables in a physical model. It formulates the statistical problem as a high-dimensional path integral and has been explored both in its exact and approximate form on chaotic and neural models~\cite{Meliza_2011,Meliza_2012,Meliza_2014,Abarbanel_2011,Ye_NaKLCa,Rey,NPG,PhysRevE2015}. Here, we show that using simulated data one can accurately estimate the parameters and forward predict all states, measured and unmeasured, in an HVC\textsubscript{RA} neuron model using this procedure. Importantly, this procedure significantly expands previous work in parameter and state estimation in that we are able to determine the time course of several unobservable variables, as well as the precise value of dozens of parameters that enter the dynamical equations {\it nonlinearly}, including parameters describing the gating kinetics~\cite{Vanier,Keren,Buhry}.

We begin with a discussion of the qualitative features of the model two-compartment neuron, in which a somatic compartment houses fast Na and K spikes, while the dendritic compartment contains slower Ca dynamics, illustrating how this model reproduces many observed features of HVC\textsubscript{RA} neurons. We then show that successful estimation of all unknown variables and parameters may be carried out to excellent accuracy via the proposed data assimilation protocol. Lastly, we discuss the role of model errors, model degeneracies, and applicability to data from HVC\textsubscript{RA} neurons in slice preparations.

In this work we do not incorporate real data; rather we perform numerical simulations known as twin experiments. In these simulations we create a set of time series for each model variable $x_a(t);\;a=1,2,...,D$ by numerically integrating the model equations. We then add noise to a small subset of these time traces; this constitutes the sparse ``data" one normally receives from neurophysiologists. These ``data," along with the model equations, are then incorporated into the assimilation procedure from which an estimate of all unknown parameters and all state variables is obtained. 

Twin experiments serve as a stress test for the data assimilation procedure: we know the actual parameter values and state variable time courses, and so can compare those known values with the estimations. They are particularly valuable to neurophysiologists in that they inform the design of laboratory experiments by determining {\it which} stimulus waveforms and {\it how many} observations are required to estimate experimentally unmeasured model states and parameters. 

This paper has companions. One will discuss the biophysics and data assimilation for models of HVC interneurons . Much less is known about the electrophysiology of these HVC components, yet we know they play a key role in the functioning of HVC~\cite{GGA1,Long_2015}, so we will devote some time to their structure and how laboratory experiments could uncover that. The broader goal is to incorporate what knowledge we have of the individual neurons in HVC to propose and test models of their connectivity in the functional HVC circuit.

%%%%%%%%%%% 			Methods 			%%%%%%%%%%

\section{Methods}
\subsection{HVC\textsubscript{RA} Neuron Model}
\label{Methods:Neuron_Model}

We model the HVC\textsubscript{RA} neuron as a Hodgkin-Huxley (HH) neuron with appropriate currents: 

\begin{align}
C_m\frac{dV(t)}{dt} &= \sum I_{ion}(t) + \sum I_{syn}(t) + \sum I_{injected}(t)
\end{align}
where $V(t)$ is the membrane voltage, $C_m$ is the membrance capacitance, $I_{syn}(t)$ are the synaptic currents, $I_{ion}(t)$ are intrinsic ionic currents, and $I_{injected}(t)$ is a specified external stimulus. The ionic currents have the familiar HH form $I_{ion}(t) = g_im_i(t)^k h_i(t)^l (E_i - V(t))$. $E_i$ is the reversal potential for the channel, $g_i$ is the maximal conductance of the channel, $m_i(t)$ and $h_i(t)$ are gating variables that describe the opening and closing of the ionic channel based on the membrane voltage, and $k$ and $l$ are integers. As we discuss isolated cells, $I_{syn}(t) = 0$.

The nonlinear voltage dependence of ionic channels in HH neurons can produce a rich variety of spiking behavior, including spike-rate-adaptations, spike rebounds, bursting, and broad depolarizations in response to simple synaptic or injected 
currents~\cite{Dayan,Ermentrout,Izhikevich}. Such behaviors can be intrinsic to the neuron, rather than arising through complex connectivity within the network. This permits stereotypy and robustness of neuron response, particularly in the presence of noise and network variability across populations within the same species. HVC\textsubscript{RA} neurons in particular are known to exhibit sparse, short bursts during vocalization~\cite{Hahnloser_2002,Fee_Book}. One could envision patterns of synaptic connections that would produce sparse bursts as are observed experimentally, but this behavior is far more robust to changes in synaptic strengths and connective variability when the bursts are produced instead by specific combinations of ionic channels in the neurons themselves.

Following experimental evidence and previous modeling efforts~\cite{Meliza_2014,Jin_2007,Long_2010,Daou}, our model contains HH K$^{+}$ and Na$^{+}$ currents that produce fast spiking in response to injected or synaptic currents and a slow, calcium-gated channel responsible for spike termination. The latter is represented by a potassium channel activated by increased intracellular calcium concentrations, $I_{K/Ca}$, with the concentration itself driven by the opening of an L-type voltage-gated high-threshold calcium channel, $I_{Ca-L}$, and decreased by a slow decay back to a background concentration. Though pharmalogical data points to the possible existence of other channels in HVC\textsubscript{RA} neurons, such as T-type Ca, A-, and Na-dependent K currents, among others~\cite{Daou,Kubota,Mooney}, we begin with this subset which appears in in the work reported below to be largely responsible for the defining long depolarizations that eventually lead to burst excitability, as well as for the short bursts themselves~\cite{Jin_2007,Long_2010}.

Our model has somatic and dendritic compartments, connected ohmically, to enhance the robustness and stereotypy of the bursts in response to injected currents amplitudes and waveforms~\cite{Jin_2007,Long_2010}. $I_{Ca-L}$ is given by a Goldman-Hodgkin-Katz form, appropriate for ions with highly disparate extracellular and intracellular concentrations~\cite{Scotland,Dayan}. As it appears that most Ca$^{2+}$ and Ca-gated channels are located in the dendritic tree~\cite{Kandel}, we locate the Ca-L and K/Ca currents in the dendritic compartment and the usual Na and K currents in the soma. 

The dynamical equations defining this model have the following form:
\\[20pt]
{\it Somatic compartment}:
\begin{align}
C_m\frac{dV_s(t)}{dt} &= g_L(E_L - V_s(t)) \nonumber\\
 &+g_{Na}m(t)^3 h(t)(E_{Na} - V_s(t))\nonumber\\
 &+ g_Kn(t)^4(E_K - V_s(t)) \nonumber\\
 &+ g_{SD}(V_d(t) - V_s(t)) + I_{inj,s}(t) 
\label{eq:dynamical_eqs_1} 
\end{align}
\\[5pt]
{\it Dendritic compartment}:
\begin{align}
C_m\frac{dV_d(t)}{dt} &= I_{Ca-L}(t) + I_{K/Ca}(t)  \nonumber\\
&+ g_{SD}(V_s(t) - V_d(t))+ I_{inj,d}(t) 
\label{eq:dynamical_eqs_2} 
\end{align}
\\[5pt]
{\it Dendritic Ca dynamics}: 
\begin{align}
&\frac{d[\text{Ca}](t)}{dt} = \phi I_{Ca-L}(t) + \frac{C\textsubscript{0}-[\text{Ca}](t)}{\tau_{Ca}} 
\label{eq:dynamical_eqs_3} 
\end{align}
\\[5pt]
{\it Gating variables}: 
\begin{align}
&\frac{di(t)}{dt} = \frac{i_\infty(V(t)) - i(t)}{\tau_i(V(t))} 
\label{eq:dynamical_eqs_4} 
\end{align}
\\[10pt]
Here, $i(t)$ is any one of the gating variables $\{n(t),\allowbreak   m(t), \allowbreak  h(t),\allowbreak   r(t)\}$, $V_s(t)$ is the membrane voltage of the somatic compartment,  $V_d(t)$ is the membrane voltage of the dendritic compartment, and we have defined the following quantities:

\bea
i_\infty(V(t)) =&& \frac{1}{2}\left(1+ \tanh\left(\frac{V(t)-\theta_i}{2\sigma_i}\right)\right) \nonumber \\
 \tau_i(V(t)) = && \tau_{i0} + \tau_{i1}\left(1-\tanh^2\left(\frac{V(t)-\theta_{\tau,i}}{2\sigma_{\tau,i}}\right)\right) \nonumber\\
+ &&\frac{\tau_{i2}}{2}\left(1+\tanh\left(\frac{V(t)-\theta_{\tau,i}}{2\sigma_{\tau,i}}\right)\right) \nonumber \\
 I_{Ca-L}(t) =&& g_{Ca-L}r(t)^2\Phi_{GHK}(t) \nonumber \\ [7pt]
 I_{K/Ca}(t)  = &&g_{K/Ca}\frac{[\text{Ca}](t)^{\eta}}{[\text{Ca}](t)^{\eta} + k_s^{\eta}}(E_K - V_d(t)) \nonumber \\ [7pt]
\Phi_{GHK}(t) =&& V_d(t)\frac{{[\text{Ca}]}_{ext}e^{-V_d(t)/V_T} -[\text{Ca}](t)}{e^{-V_d(t)/V_T} -1}  
\label{eq:dynamical_eqs_5}
\eea
\\[10pt]
The specific form of the time constant of gating functions, $\tau(V)$, reflects the combination of previously used functional forms found in the literature~\cite{Meliza_2012,Jin_2007,Long_2010,Daou}. As we place the Na and K currents in the soma, then $i_{\infty}(V(t)) = i_{\infty}(V_s(t))$ and $\tau_i(V(t)) = \tau_i(V_s(t))$ for $i = \{n(t),m(t),h(t)\}$, while the calcium dynamics is placed in the dendrite, whereby $i_{\infty}(V(t)) = i_{\infty}(V_d(t))$ and $\tau_i(V(t)) = \tau_i(V_d(t))$ for $i = \{r(t)\}$. Finally, we note that the behavior of the neuron appears to be relatively insensitive to the values of $\tau_{i1}$,  $\tau_{i2}$, $\theta_{\tau,i}$, and $\sigma_{\tau,i}$.

\subsection{Path Integral Methods of Data Assimilation}
\subsubsection{General Procedure}

The term ``data assimilation'' arises in the geophysical literature and refers to analytical and numerical procedures in which information in measurements is transferred to model dynamical equations selected to describe the processes producing the data. The goal of assimilating the data is to estimate the fixed parameters of the model and the time course of unmeasured state variables. 

The difficulty of data assimilation arises not only from the possible complexity and high dimensionality of the system, nor just from inherent noise in the measurements, but also from the sparsity of actual observations. This sparsity is associated with physical limitations in performing the experiments themselves. In a neuronal system, for example, direct measurement of the time course of the gating variables and most ionic concentrations is not now possible. This leaves the subset of measurable state variables at only one out of several, namely the membrane voltage at the soma. The unobserved state variables are coupled to the measured variables through the model equations and may therefore be ascertained from the data. Nevertheless, the limitations of available numerical methods, paucity of measured components, and possible degeneracies of the system often render accurate predictions formidable. 

%In saying that only the soma voltage can be measured directly, we reserve the possibility that high spatial resolution techniques for observing the cystolic Ca concentration [Ca]$(t)$ may be developed in the future.

We describe the problem in the following manner. The data presented to the model in response to a known external stimulus consists of noisy measurements made at times $\{t_0,t_1,...,t_m\}$ within a time period known as the estimation window, $[t_0,t_m]$. We seek to determine both the model state variables at the end of the estimation window ${\x}(t_m)$ and the unknown model parameters $\bf {p}$. This transfer of information from data to the model is the data assimilation procedure. 

Incorporating the estimated $\bf {p}$ into the model and beginning at the estimated state of the model at $t_m$, we predict the response of the system for times $t > t_m$ by integrating the dynamical equations from that point onward or, when model errors are accounted for, projecting forward in time a probability distribution for the model state. 

Data assimilation often involves at some level the minimization of a cost function, quantifying the deviation of the model output from the observed data in some estimation window. In many cases, once this optimization procedure is carried out, it is found that the trajectory of the measured variables coincides quite well with the data within in this window. However, the quality of the estimation of parameters and unobserved state variables cannot be ascertained without further tests; the true test of the assimilation procedure is comparison of the {\it predictions} of the state variables for $t > t_m$. It is often found that excellent estimations lead to unsatisfactory predictions of the measured states. In all estimations carried out in the present work, we base the validity of our estimation entirely on comparisons of the corresponding predictions.

Here, we define our problem as a path integral realization of a statistical data assimilation procedure~\cite{Abarbanel_book}. It can be shown that the conditional probability of the final state ${\bf x}(t_m)$ at the end of a measurement window, conditioned on the observations ${\Y} = \{{\bf y}(t_0)...$ ${\bf y}(t_n)...$ ${\bf y}(t_m)\}$ within that window, equals an integral of the conditional probability $P({\X}|{\Y}) $ = $\allowbreak \exp[-A_0({\X,\Y})]$ over intermediate states ${\X}_m = \{{\bf x}(t_0)...$ ${\bf x}(t_n)...$ ${\bf x}(t_m)\}$~\cite{Abarbanel_book}:

\begin{align}
P({\x}(t_m)|{\Y}) =&\int d{\bf X}\exp[-A_0({\X,\Y})] \nonumber \\ 
A_0({\X,\Y}) =& -\sum_n \log[P({\bf y}(t_n)| {\bf X}_n, {\bf Y}_{n-1}) \nonumber \\
&-\sum_n \log[P({\bf x}(t_{n+1})| {\bf x}(t_{n}))\nonumber \\
& - \log[P({\bf x}(t_0))] 
\\ \nonumber
\end{align}

The conditional probabilities incorporate the measurement uncertainty and the model dynamics, the latter of which is expressed through the state- and parameter-dependent discrete time maps (the discretized form of the dynamical equations), 

\begin{align}
x_a(t_{n+1}) &=  f_a(\x(t_n),\p)\\ \nonumber
\end{align}

Due to the analogy of this path integral formulation to that in Lagrangian dynamics, $A_0(\X)$ is referred to as the ``action". If the model errors and measurement noise are Gaussian, $A_0({\X})$ assumes a simple form~\cite{Abarbanel_book}:

\begin{align}
\label{eq:action_gaussian}
A_0({\bf X}) &= \sum_{n=0}^{m}\frac{R_m(t_n)}{2}\sum_{l=1}^L[x_l(t_n) - y_l(t_n)]^2 \nonumber \\
&+ \sum_{n=0}^{m-1}\sum_{a=1}^{D}\frac{R_{f}(a)}{2}[x_a(t_{n+1}) - f_a({\bf x}(t_{n}),{\bf p})]^2 \\ \nonumber
\end{align}

Here, $R_m$ and $R_f$ are the inverse variance of the measurement errors and model errors, respectively. $L$ is the number of measured variables; $D$ is the number of state variables, and the number of time points at which observations are made is $m+1$. The term $-\log[P(\x(t_0))]$ reflects prior knowledge of the state of the system at the onset of the estimation window. However, in a practical sense we often have no information about the distribution of the state variables when measurements begin, and therefore represent this term by a uniform distribution over the model dynamical range. This presents us with a constant in the action which can be ignored~\cite{Abarbanel_book}.
 
Previous work has utilized a Metropolis-Hastings Monte Carlo (MHMC) method for evaluating the path integral, and has been successful in various models with acceptable computational efficiency~\cite{Meliza_2012}. But while the MHMC method has the advantage that it can be implemented computationally in a parallel architecture, evaluating the full path integral can prove a formidable numerical challenge for most physically practical problems. In the spirit of Lagrangian dynamics and perturbative field theory, one may more systematically approach the problem by expanding around stationary paths, that is, utilizing Laplace's method~\cite{laplace,Zinn}. One may then work to higher orders in this expansion to the accuracy desired. 

The computational difficulty of the problem is thus shifted to one of nonlinear optimization -- to finding the minima of $A_0({\bf X})$ \cite{Abarbanel_book}. The form of the action given in Eq. (\ref{eq:action_gaussian}) is identical to the cost function used in optimization processes known widely in the geophysics community as ``Weak 4DVar", as the model constraints are not enforced strictly, but in proportion to the magnitude of $R_f$. This indicates that the 4DVar approach is in fact an approximation to the path integral formulation of the statistical problem, to which higher order corrections can be systematically calculated \cite{Abarbanel_book}.

\subsubsection{Annealing in a Variational Approximation of the Path Integral}

The fact that the dynamical equations of biological neuron models are highly nonlinear implies that the action can be nonconvex, exhibiting many local minima. In Laplace's approximation, the global minimum is the most relevant, contributing most strongly to the path integral; however, it is difficult to both detect such a minimum and prove it is the lowest. Here, we will employ an extension to Laplace's approximation to the data assimilation path integral, incorporating an iterative annealing-like procedure in which $R_f$ is not held fixed, but changed successively. This method has been shown to be effective in state and parameter estimation in archetypal chaotic models such as the Lorenz96 model \cite{NPG} as well as in simple neuron models. Here, the nonlinear optimization is first carried out for a relatively small (possibly zero) value of $R_f, R_{f0}$. As $R_{f0}$ is small, the model constraints at the first step of the annealing procedure are enforced weakly if at all. The set of all solutions is thus quite degenerate, consisting of all paths whose measured components match the data and whose unmeasured components are unrestricted. 

\begin{figure}[htbp] % float placement: (h)ere, page (t)op, page (b)ottom, other (p)age
 \centering
\includegraphics[width=0.5\textwidth]{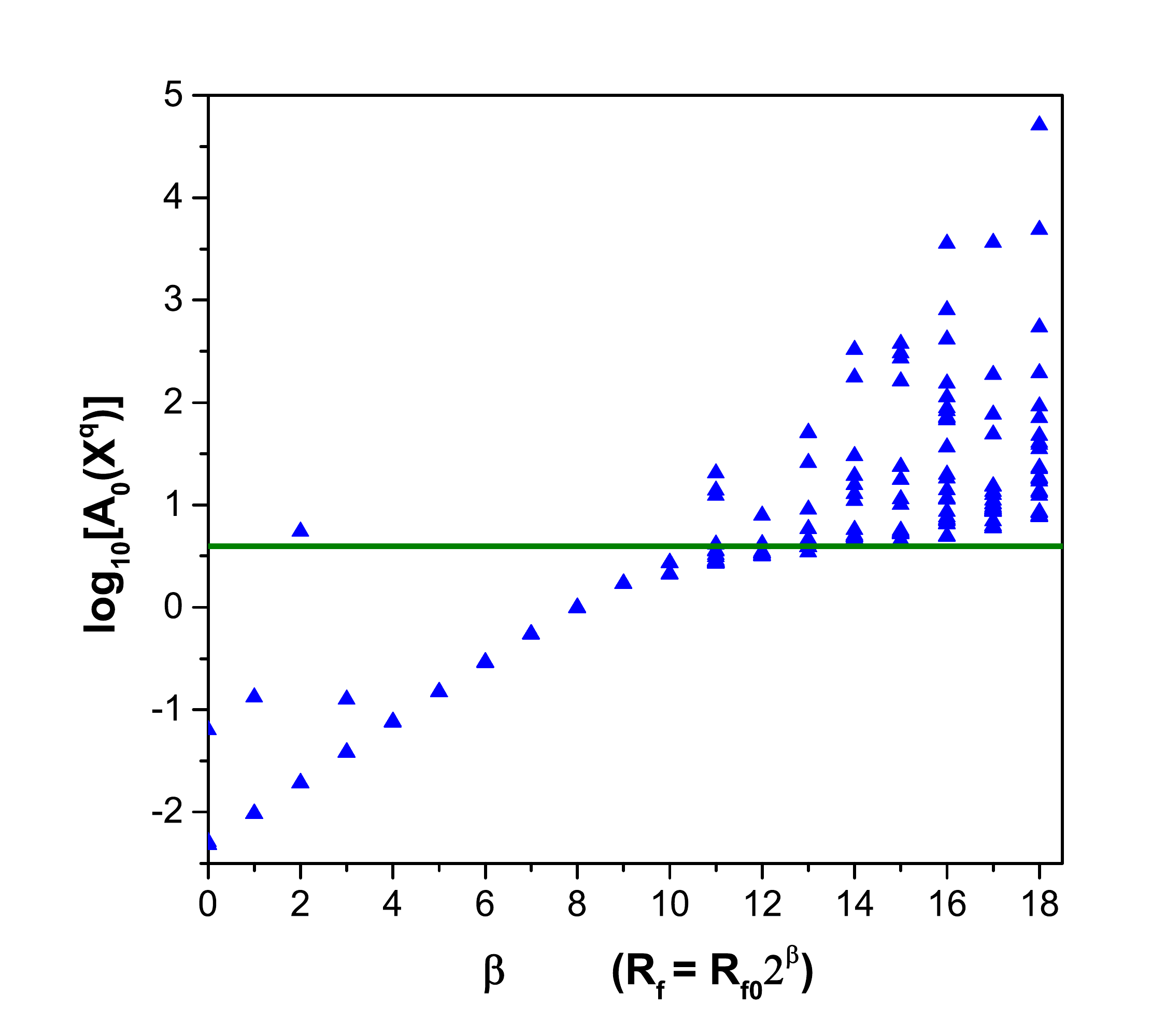}
  \caption{An illustrative action level plot for the annealing method of statistical data assimilation. Many degenerate action levels at low $\beta$ split as $\beta$ is increased, indicating the removal of degeneracies of local minima. In this example, the lowest minimum is at $\log_{10}(A_0(\X^0)) \approx 0.6$, which is indicated by the heavy horizontal line. The calculation presented is for the full HVC\textsubscript{RA} model. It shows that the lowest action level does not split off substantially from the other levels and does not go to the consistent lowest minimum action level. This feature led us to seek a different protocol for estimating the fast and slow dynamical variables separately}
  \label{figure1actionplot}
\end{figure}

The result of the optimization is then used as the initial guess for a subsequent optimization of the same $A_0$, but with $R_{f0}$ increased by some factor, $\alpha^{\beta};\;\alpha > 1$. This procedure is then repeated for a number of steps, the $(m+1)\cdot D$-dimensional action surface changing as the model equations are more heavily enforced with increasing $\beta$. If several of these procedures are carried out in parallel, then as $\beta$ is increased, the degenerate surface for $R_f \approx 0$ splits into various local minima that may or may not individually coincide with the lowest minimum (see Fig. \ref{figure1actionplot}). 

The advantages of this annealing procedure are discussed in detail elsewhere~\cite{NPG}; here we note only that a) the procedure confronts the problem of ill-conditioning of the Hessian provided by strong model constraints, i.e. when $R_f$ is large, and b) the existence of a consistent lowest minimum of the action, and therefore the consistency of the model with the data, can be checked by comparing the limiting action value with~\cite{NPG}:
\bea
\nonumber \\
A_0(\X^0) = \frac{R_m \sigma^2}{2}L(m+1),  \\ \nonumber 
\eea
where $\sigma^2$ is the variance of the measurement noise. This value is indicated by the horizontal line in Fig. \ref{figure1actionplot}. A comparison of this value with those in the action level plots at high $\beta$ readily allows identification of incongruities between the model equations and the measured data.

%%%%%%			Figure 1

In our work optimization was performed using the interior-point algorithm provided by the open source software IPOPT \cite{Wachter}, utilizing the ma57 linear solver library, on a standard desktop computer. The dynamical equations were discretized in the action using a Runge-Kutta 4th order approximation. Data was sampled at 0.02 ms, the same as the timestep of the discretization of the model, for 600 ms, or 30001 timepoints. Hard bounds on the parameters and state variables during the optimization must also be supplied to IPOPT; the bounds for the voltages are between -120 to 50 mV, intracelllular calcium concentration between 0 and 10.0 $\mu$M, and the gating variables between 0 and 1. 

\subsubsection{Twin Experiments in Data Assimilation}

%%%%  				Figure 2

\begin{figure*}[htbp] % float placement: (h)ere, page (t)op, page (b)ottom, other (p)age
  \centering
  \includegraphics[width=.83\textwidth]{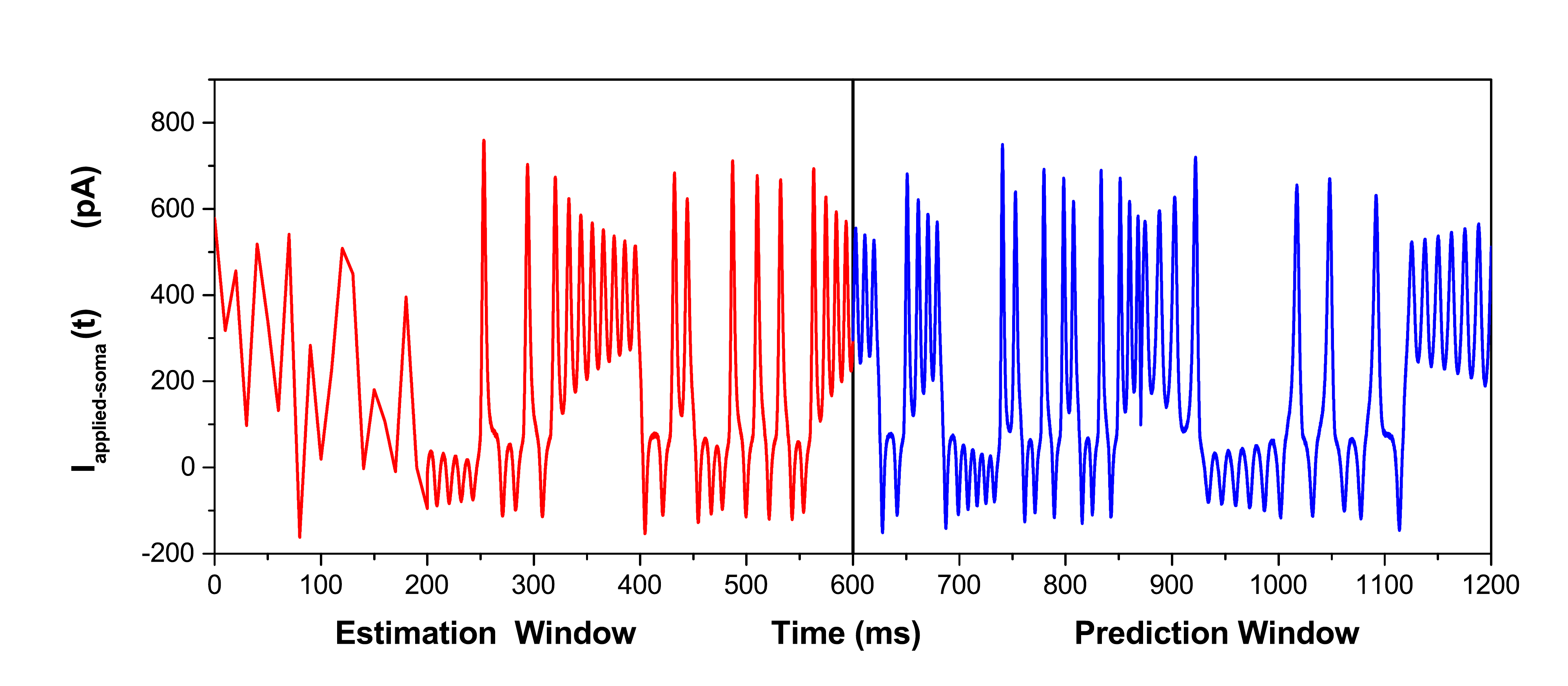}
  \caption{The applied soma current waveform used in the data assimilation procedures is a combination of pseudo-noisy currents and the output of a chaotic model. The pseudo-noisy currents, for the time segment $0 \mbox{ ms} \le t \le 200 \mbox{ ms}$, are created by uniformly sampling current values between -200 and 600 pA every $10 \mbox{ ms}$ and linearly interpolating between those times. For $200 \mbox{ ms} \le t \le 600 \mbox{ ms}$, a waveform produced by the output of the chaotic Lorenz63 model is used as the stimulating current. The first 600 ms is used in the data assimilation routine to estimate the parameters and time courses of the state variables. The last 600 ms of stimulating current, also from the Lorenz63 model, is then used within the prediction window where the validity and accuracy of the parameter and state estimation is ascertained}
  \label{iappliedsoma}
\end{figure*}

Due to the difficulties and complexities involved in successful data assimilation of high-dimensional, chaotic systems, it is often instructive to perform ``twin experiments" before handling actual data. In such simulations, ``data" is generated from the model equations themselves \cite{Abarbanel_book}. A subset of these solutions to the model is then chosen and noise is added to them to act as the ``measured data" to be used in the action. Since the data is generated by simulation, rather than extracted from experiments, the true trajectories of both measured and unmeasured variables are already known. One can unambiguously test the performance of the data assimilation procedure by comparing estimations and predictions of {\it all} variables and parameters. While twin experiments do not provide incontrovertible proof that data assimilation will be successful for the given combination of model equations and actual measured data, they nevertheless give a strong indication that the data set used in the assimilation is sufficiently complete to accurately estimate the unknown states and parameters: this is the central question of data assimilation.

Traditional currents such as steps, saws, and pulses are unable to adequately sample the full phase space of the system. As such, we use complex currents to generate the ``data" used in the twin experiments. In particular, we use a somatically injected current whose waveform is generated from a combination of pseudo-noisy uniform sampling and the output of the Lorenz63 system in its chaotic regime (Fig. \ref{iappliedsoma}). Following anticipated experimental limitations, we treat soma membrane voltage as the only measured variable. Therefore, this voltage trajectory, to which Gaussian noise is added at each time point, is the only measurement presented to the action. The assimilation routine is then carried out and the results of the estimated parameters and unobserved states are compared to the true trajectories determined from the forward integration of the known model.

%%%%%%%%%%%				RESULTS 				%%%%%%%%%%%%

\section{Results}
\subsection{Qualitative Behavior of HVC\textsubscript{RA} Model}
\label{sec: qual_behavior}

%%%% 			  Table 1

\begin{table}
%\centering
\begin{tabular}{ c  c  c |c  c  c }
%Param. & Value & Units & Param. & Value & Units \\
\hline\noalign{\smallskip}
$g_K$ & 120 & nS &  $g_L$ & 3 & nS\\
$g_{Na} $ & 1050&  nS & $g_{SD} $ & 5 &nS \\
$g_{Ca-L}$ & 0.06 &nS/$\mu$M & $g_{Ca/K}$ & 240& nS \\
 $E_K$ & -90& mV &  $E_{L}$ & -80& mV  \\
$ E_{Na}$ & 55 & mV &  $V_T$  &13.5& mV \\
$\theta_n$ & -35& mV & $\sigma_n$ & 10 &mV \\
 $\sigma_{\tau,n}$& -15& mV & $\tau_{n0}$ & .1& ms  \\
$\theta_{\tau,n}$ & -27& mV  &$\tau_{n2}$ & .5 & ms \\
$\theta_m$& -30  & mV & $\sigma_m$& 9.5 &mV \\
 $\sigma_{\tau,m}$& 0 &mV &    $\tau_{m0}$ & .01& ms \\
$\theta_{\tau,m}$ & 0& mV  & $\tau_{m2}$ & 0 & ms\\
$\theta_h$ & -45  & mV & $\sigma_h$ & -7& mV  \\
$ \theta_{\tau,h}$ &-40.5& mV &  $\tau_{h0}$ & .1& ms  \\
 $\sigma_{\tau,h}$& -6&  mV  &  $\tau_{h2}$ & .75 &ms \\
$\theta_r$& -40  & mV & $\sigma_r$& 10 &mV \\
 $\sigma_{\tau,r}$ &0 & mV  &  $\tau_{r0}$ & 1 &ms \\
$ \theta_{\tau,r}$ &0& mV & $\tau_{r2}$ & 0 &ms \\
 $\phi $ & 8.67e-5& $\mu$M/pA/ms  & $ \tau_{Ca}  $& 33 &ms  \\
$\eta $& 2 &---& $C_{0}$ & 0.48& $\mu$M \\
 $C_m$& 21 & pF & $k_s $ & 3.5& $\mu$M \\
$[\text{Ca}]$\textsubscript{ext}  & 2500 &$\mu$M \\
\noalign{\smallskip}\hline
%\hline\noalign{\smallskip}
\end{tabular}
\caption{Parameter values for the two-compartment model with tuned calcium dynamics and inter-compartment coupling. Parameters listed in Eqs. (\ref{eq:dynamical_eqs_1}) - (\ref{eq:dynamical_eqs_5}) which are not shown here are set to zero }
\label{tab:qualitative_behavior}
\end{table}

%%%% 			  Figure 3 

\begin{figure}[htbp] % float placement: (h)ere, page (t)op, page (b)ottom, other (p)age
  \centering
  \includegraphics[width=.52\textwidth]{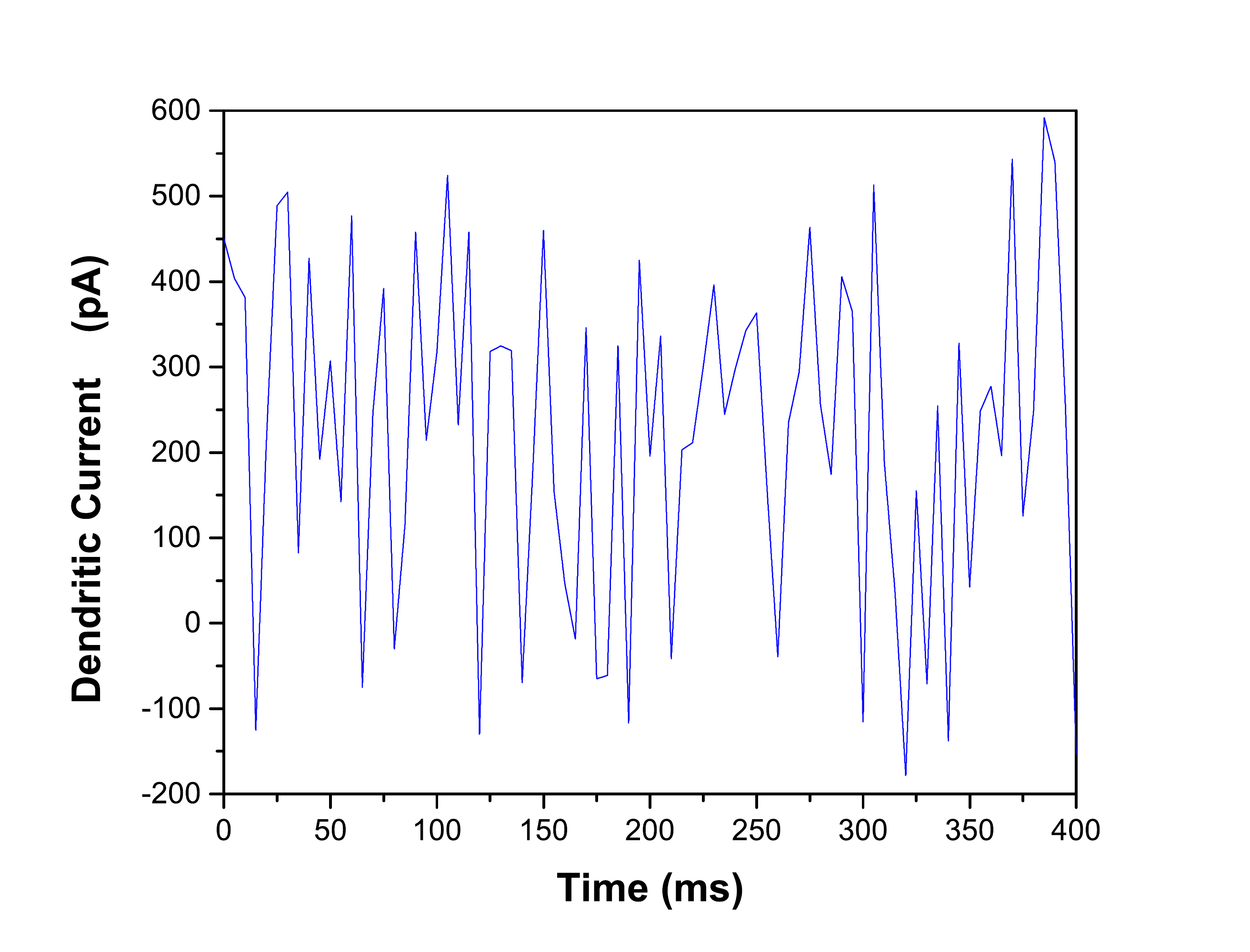} \\
  \includegraphics[width=.52\textwidth]{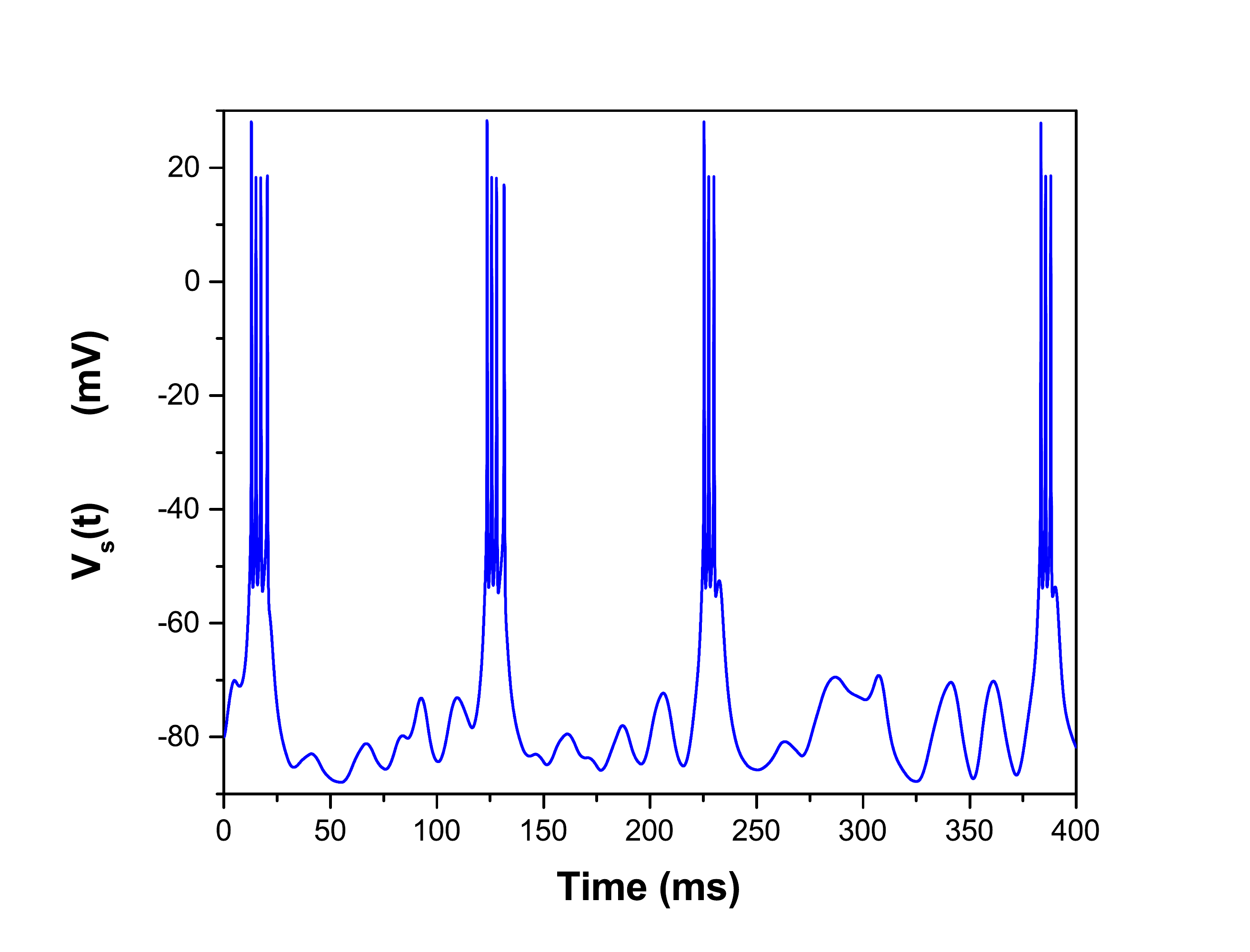} \\
  \caption{The two-compartment HVC\textsubscript{RA} model reproduces stereotyped bursts in response to pseudo-noisy dendritic currents. The currents are created by uniformly sampling values between -200 and 600 pA every 5 ms, and linearly interpolating between (upper graph). Varying the temporal resolution of the noise from 5 ms to 10, 20, or 40 ms changes neither the average duration of the burst nor the average spike frequency within the burst (lower graph)}
  \label{hvcraburstsisample5ms}
\end{figure}

With a particular choice of parameters, of which one example set is shown in Table \ref{tab:qualitative_behavior}, the proposed model reproduces many qualitative features of HVC\textsubscript{RA} neurons uncovered by experiment. The most conspicuous feature of the HVC\textsubscript{RA} neurons is the sparse, stereotyped bursting they display during vocalization and sleep \cite{Hahnloser_2002}. These bursts themselves are generic, with a relatively small variance in length and number of spikes per burst. We reproduce such bursts in Fig. \ref{hvcraburstsisample5ms}, in which we simulate the response of the neuron with respect to pseudo-noisy dendritic currents. The burst waveforms are largely insensitive to the temporal resolution and magnitude of this noisy current.

%%%%       			Figure 4

\begin{figure*}
  \centering
  \includegraphics[width=.45\textwidth]{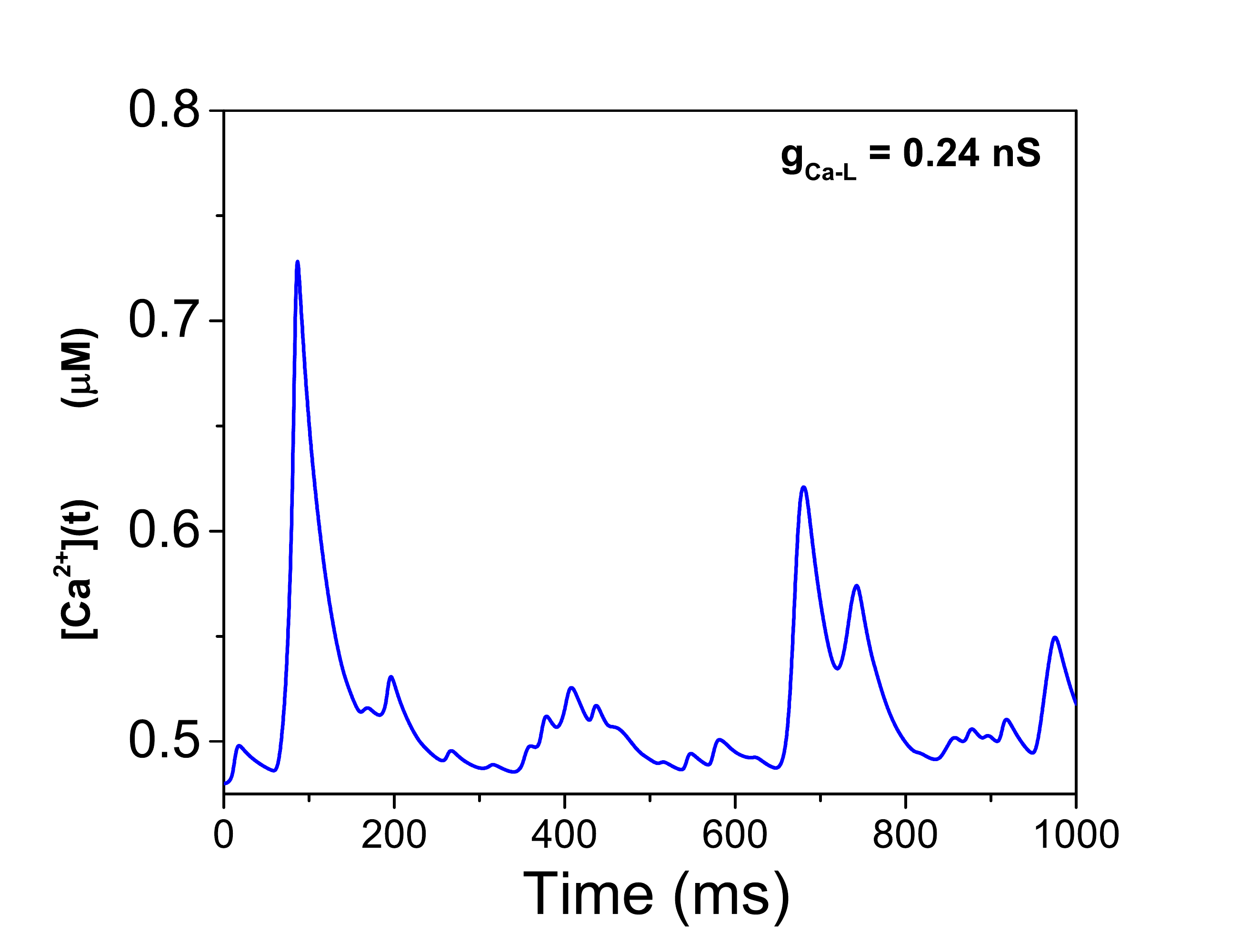} 
  \includegraphics[width=.45\textwidth]{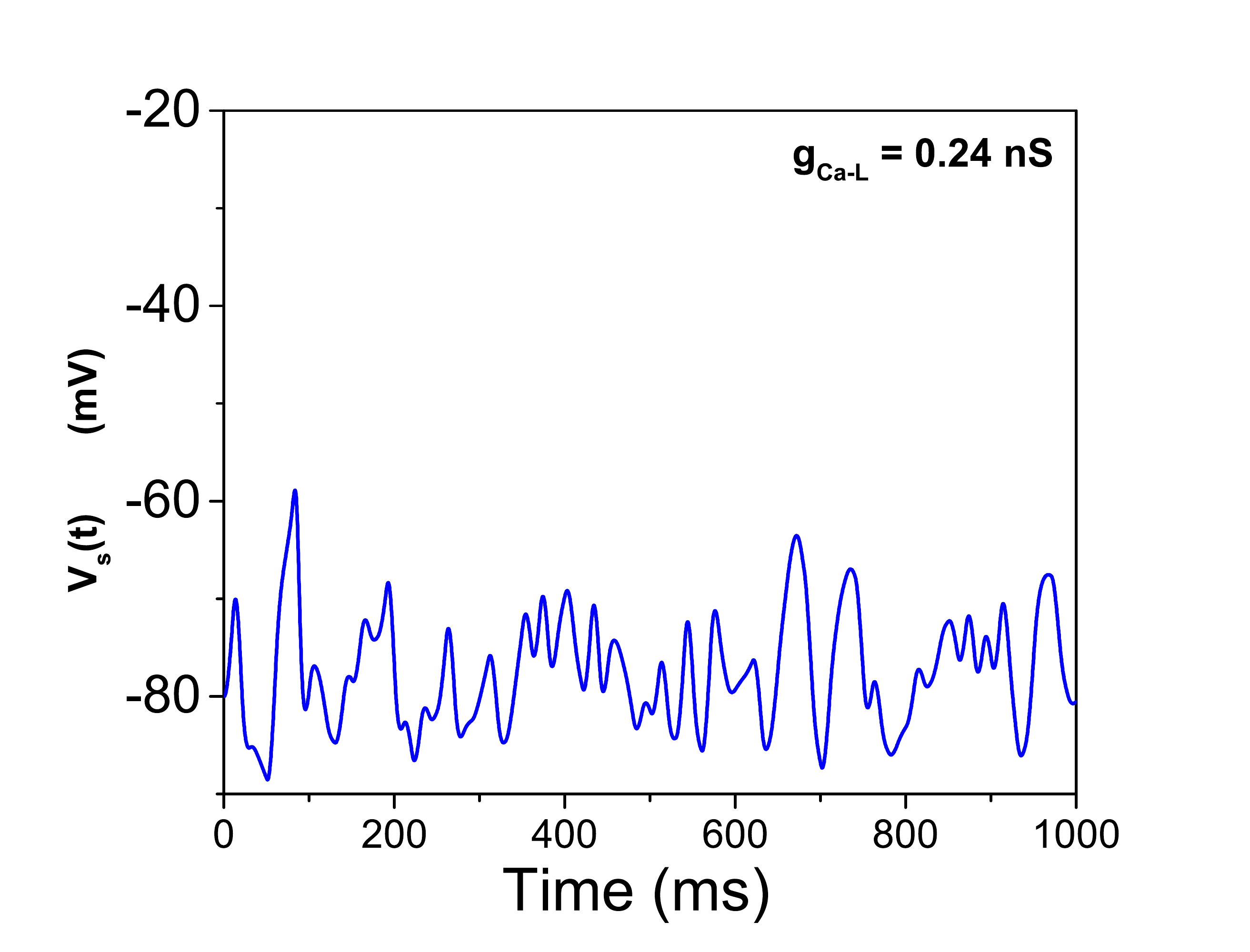} \\
  \includegraphics[width=.45\textwidth]{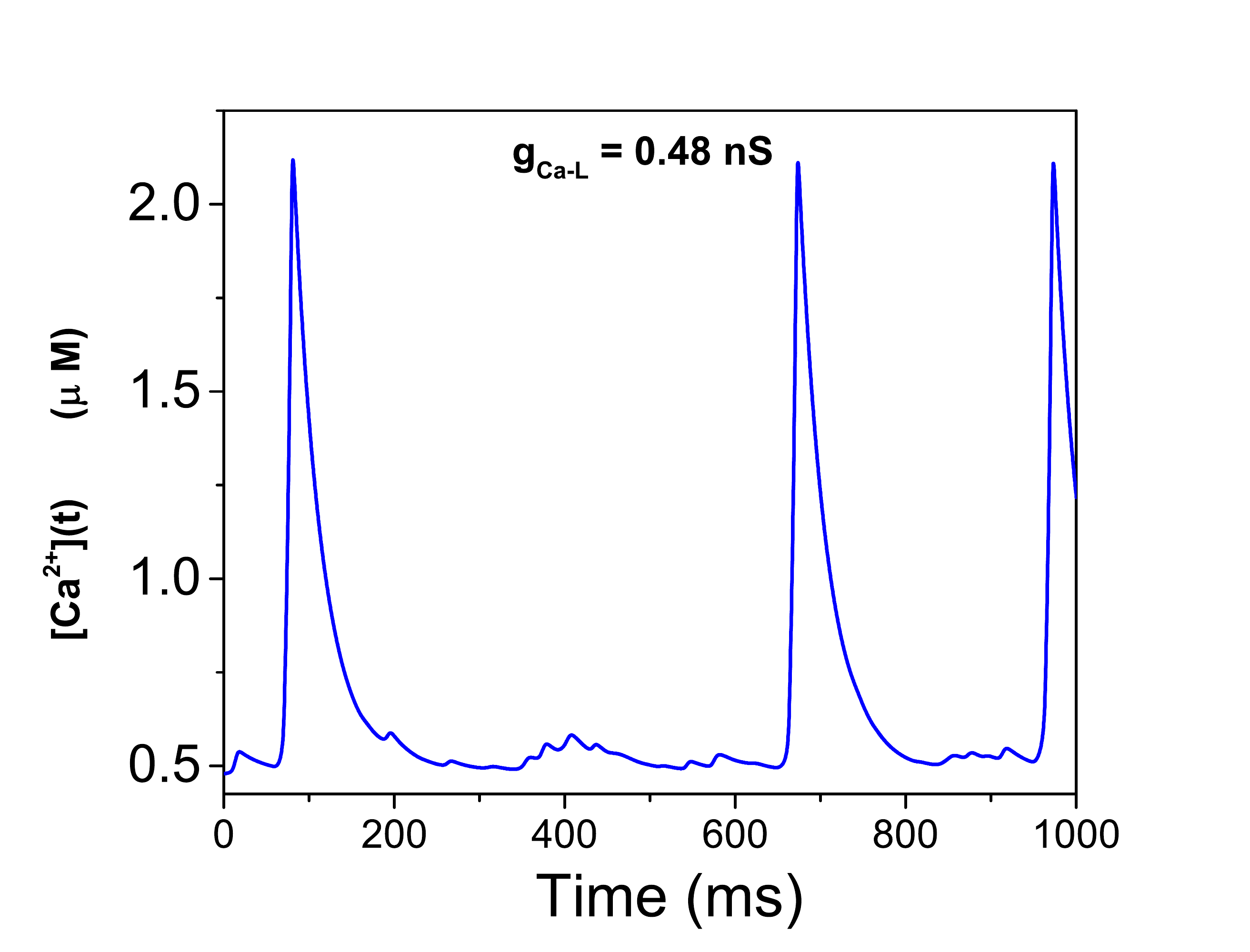} 
  \includegraphics[width=.45\textwidth]{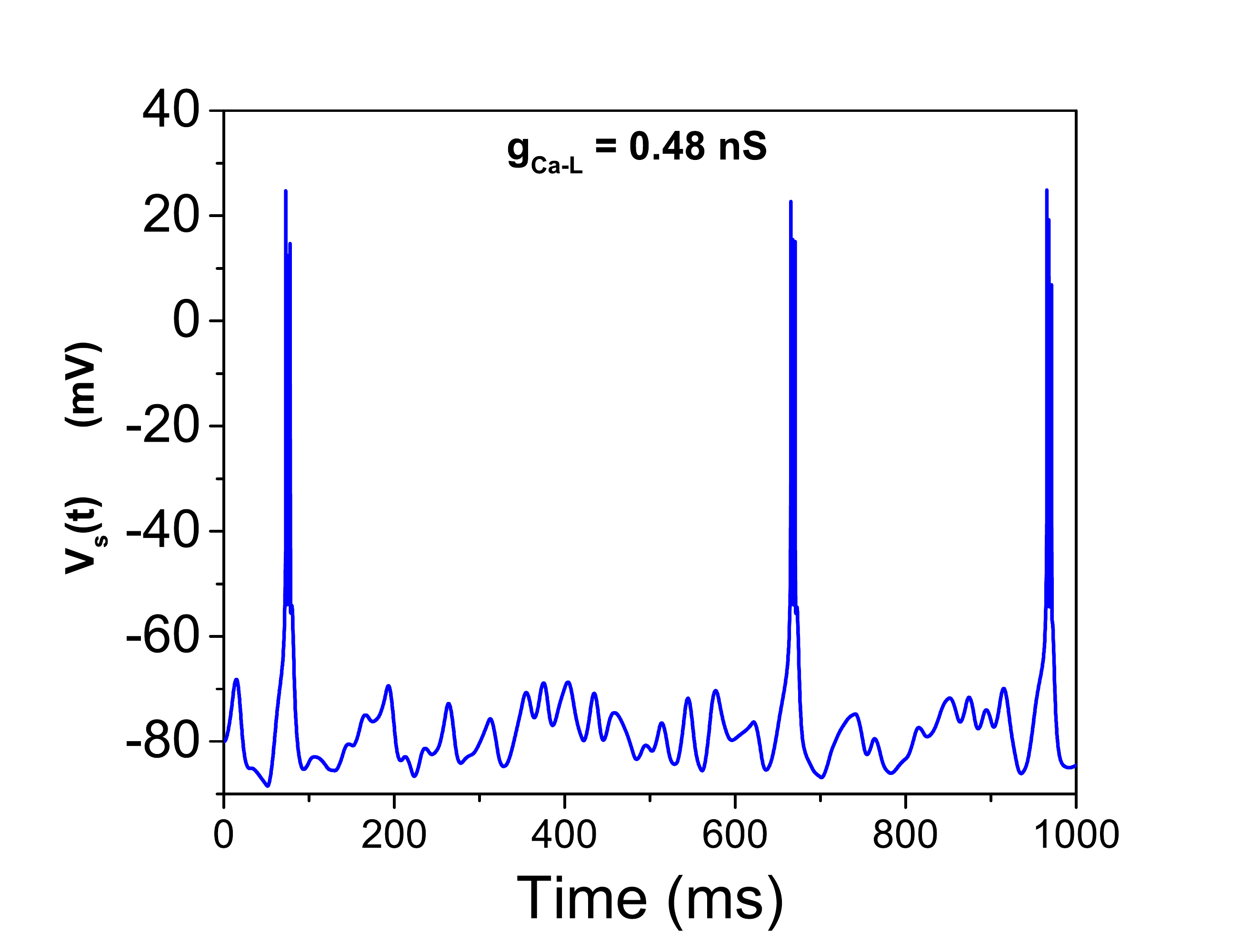} \\
  \includegraphics[width=.45\textwidth]{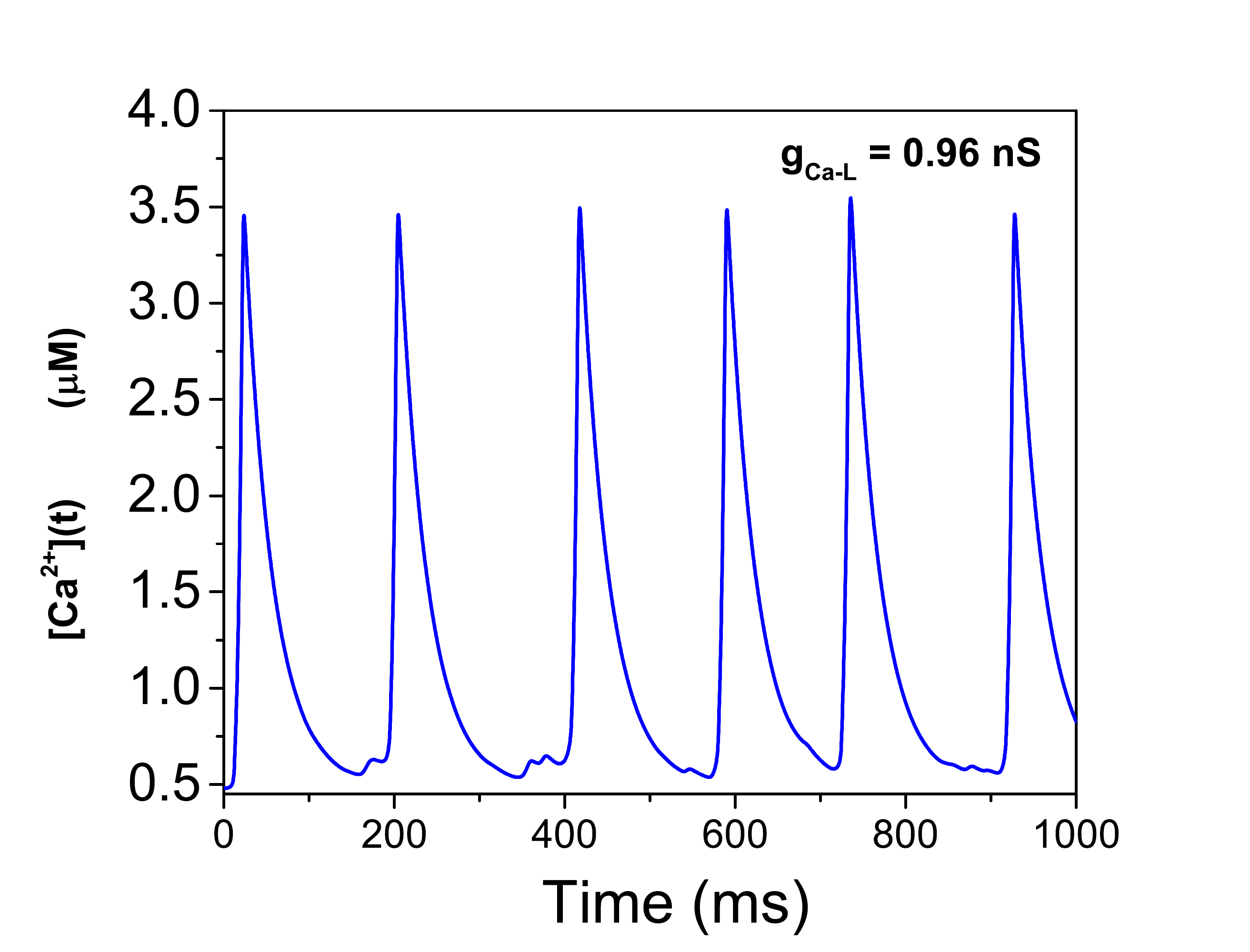} 
  \includegraphics[width=.45\textwidth]{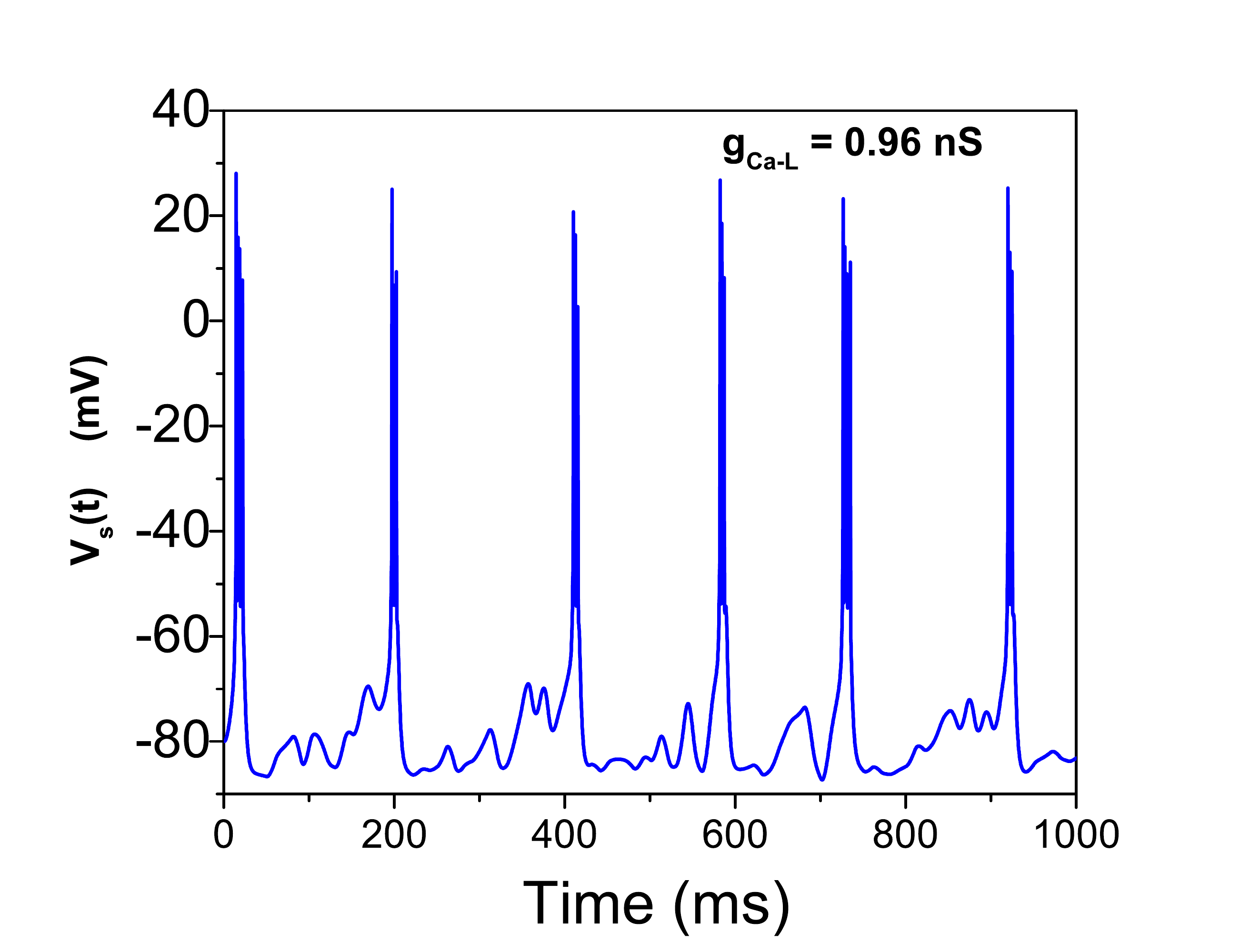}
  \caption{The effect of changing the strength of the $I_{Ca-L}$ currents on the burst incidence. As $g_{Ca-L}$ is scaled from $g_{Ca-L} = 0.24$ nS (40\% of its original value) to  $g_{Ca-L} = 0.96$ nS (160\% of its original value), somatic burst incidence increases, and inter-burst intervals decreases, in accord with experimental observations~\protect\cite{Long_2010}. Despite the increased burst incidence, the stereotypy of the bursts is preserved}
  \label{fig:Ca_agonist}
\end{figure*}

Bursting can be understood as arising from the interplay of fast and slow currents~\cite{Izhikevich}. The fast Na and K currents in the somatic compartment produce the spikes, and the slow Ca and Ca/K currents in the dendrite modulate the behavior of the soma periodically from spiking to quiescence. Bursting, as opposed to tonic spiking, occurs because the slow Ca concentration modulates the system slowly between spiking and resting. In particular, at the onset of bursting, injected dendritic currents open calcium channels, eliciting a dendritic voltage spike. As the dendrite voltage is high, its coupling via $g_{SD}$ therefore effects fast spiking in the soma: this is the burst. The spiking then terminates because the now raised [Ca] concentration opens the outward K/Ca current, now driving the dendritic voltage back to baseline. The key point is that slowly moving calcium takes time to leave the cell, so while the dendritic voltage has been terminated, the K/Ca channel is not yet fully closed. This suppresses somatic spiking for some time after the burst. Only when [Ca] has returned to a sufficiently low level does this channel close, then allowing the process to repeat. 

Bursting is thus caused by appropriate combinations of the parameters in the calcium current, the K/Ca current, and the calcium dynamics. In fact, as we will see later, this combination of parameters may not be unique.

Next, we investigated the effect of calcium channel changes. It has been reported that calcium channel enhancers and blockers can lead to changes in burst incidence during sleeping; in particular, the presence of antagonists causes lower burst incidence and longer inter-burst intervals, while the presence of agonists brings about the opposite effect~\cite{Long_2010}. We simulate this by comparing the somatic voltage trace again in response to the same noisy dendritic currents, with $g_{Ca-L}$ is varied. As shown in Fig. \ref{fig:Ca_agonist}, as $g_{Ca-L}$ is decreased, calcium-mediated depolarizing events are less likely to bring about bursts, resulting in longer inter-burst intervals.

%%%%% 			Figure 5 

\begin{figure*}
  \centering
  \includegraphics[width=.49\textwidth]{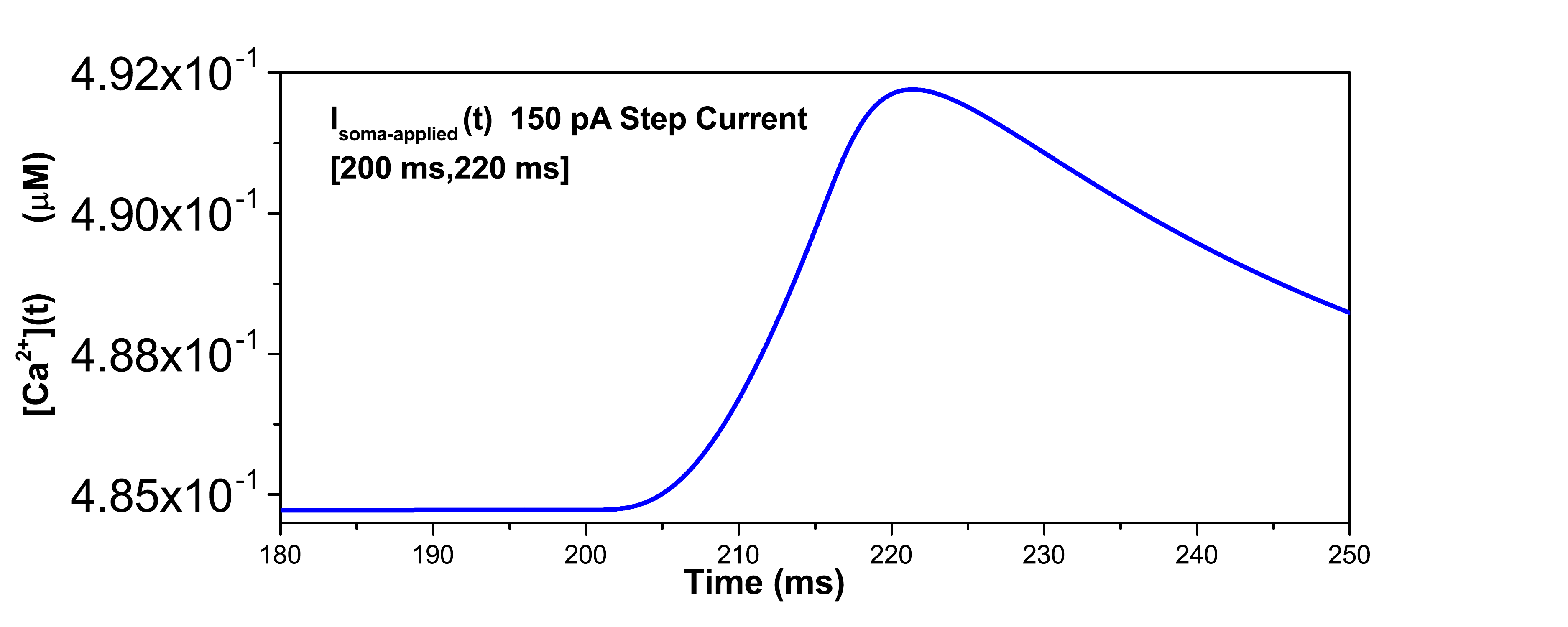} 
  \includegraphics[width=.49\textwidth]{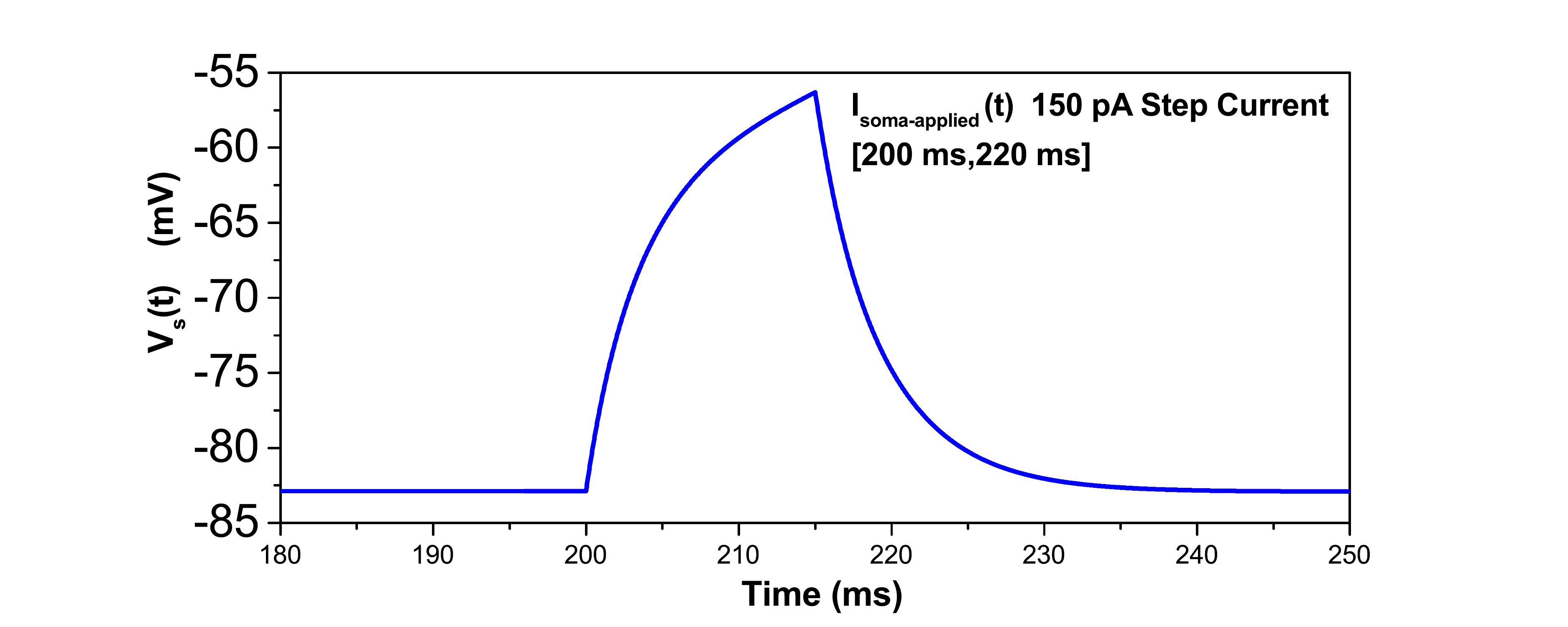} \\
  \includegraphics[width=.49\textwidth]{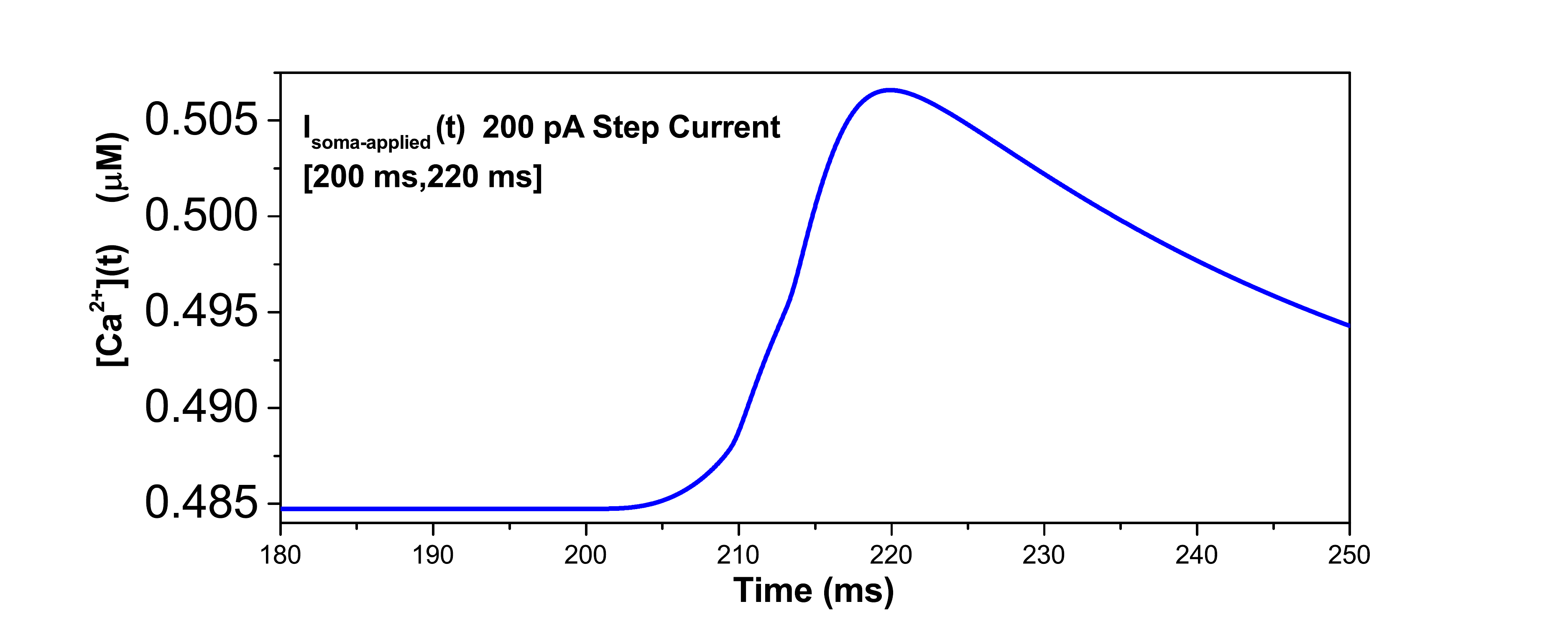} 
  \includegraphics[width=.49\textwidth]{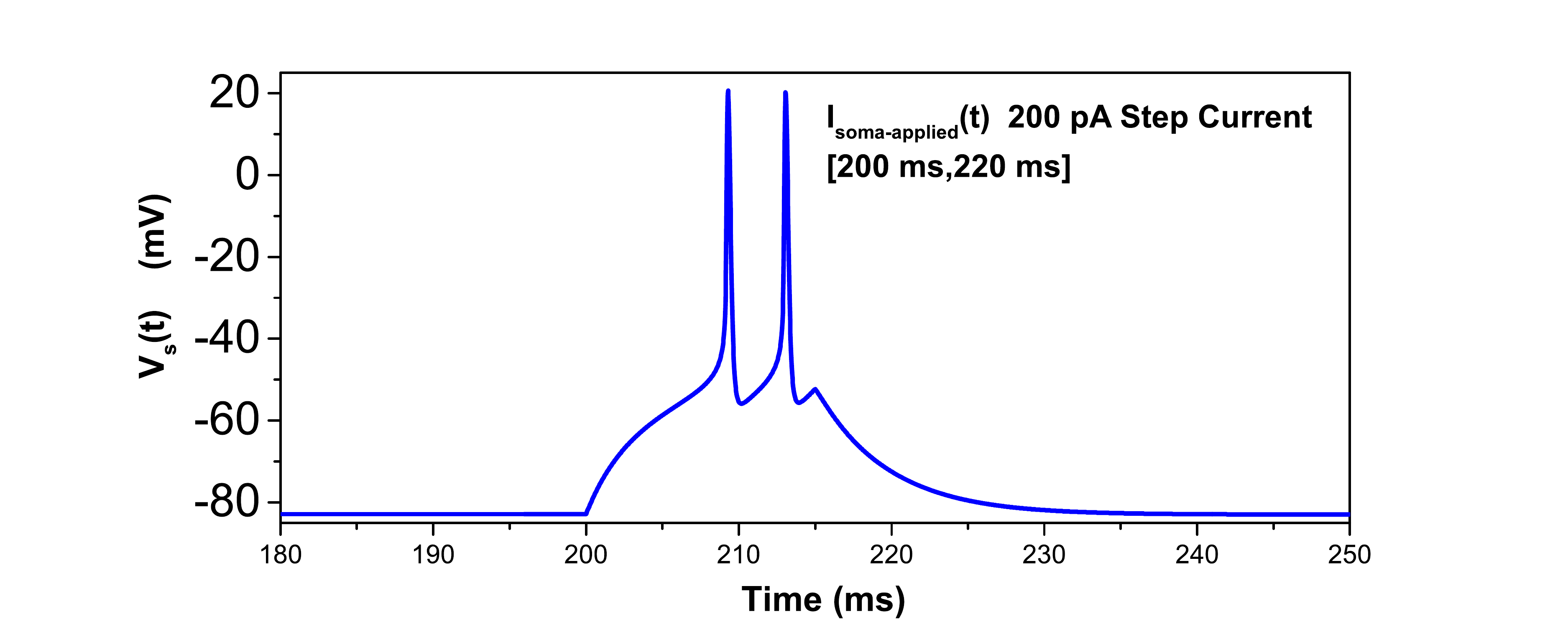} \\
  \includegraphics[width=.49\textwidth]{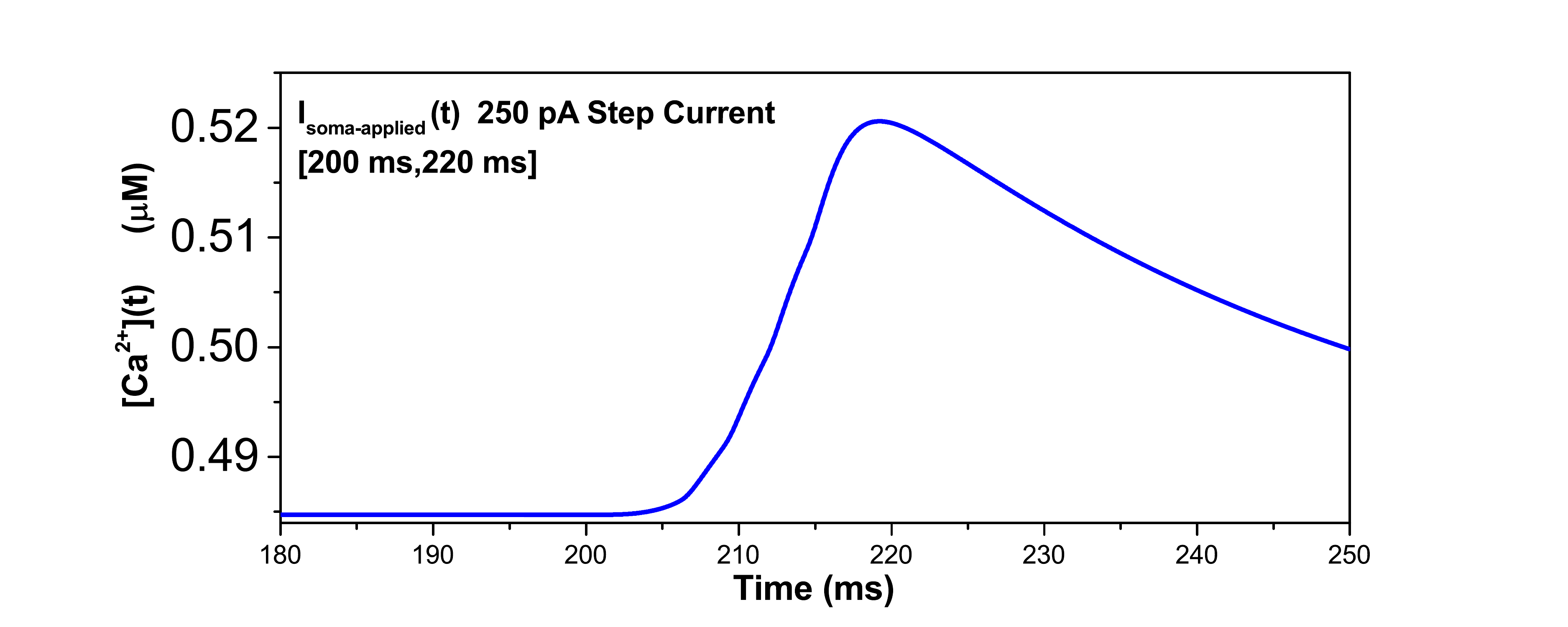} 
  \includegraphics[width=.49\textwidth]{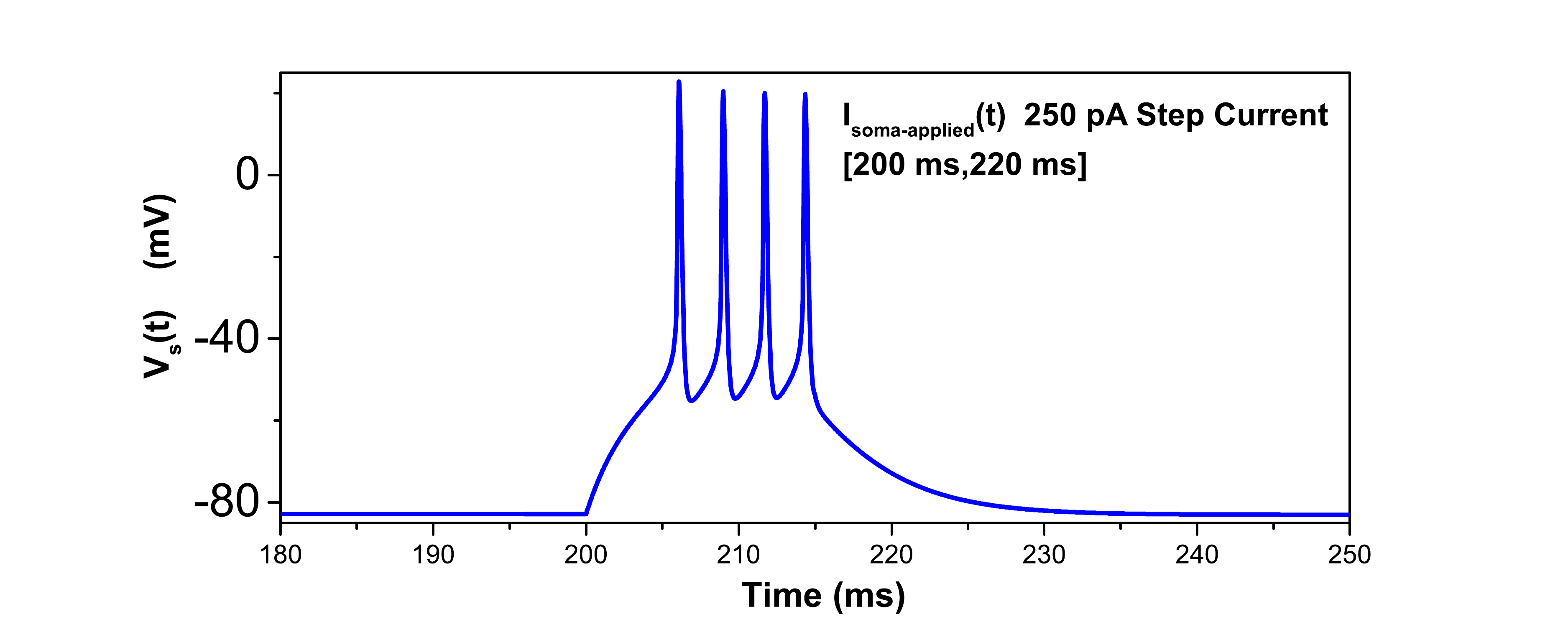} \\
  \includegraphics[width=.49\textwidth]{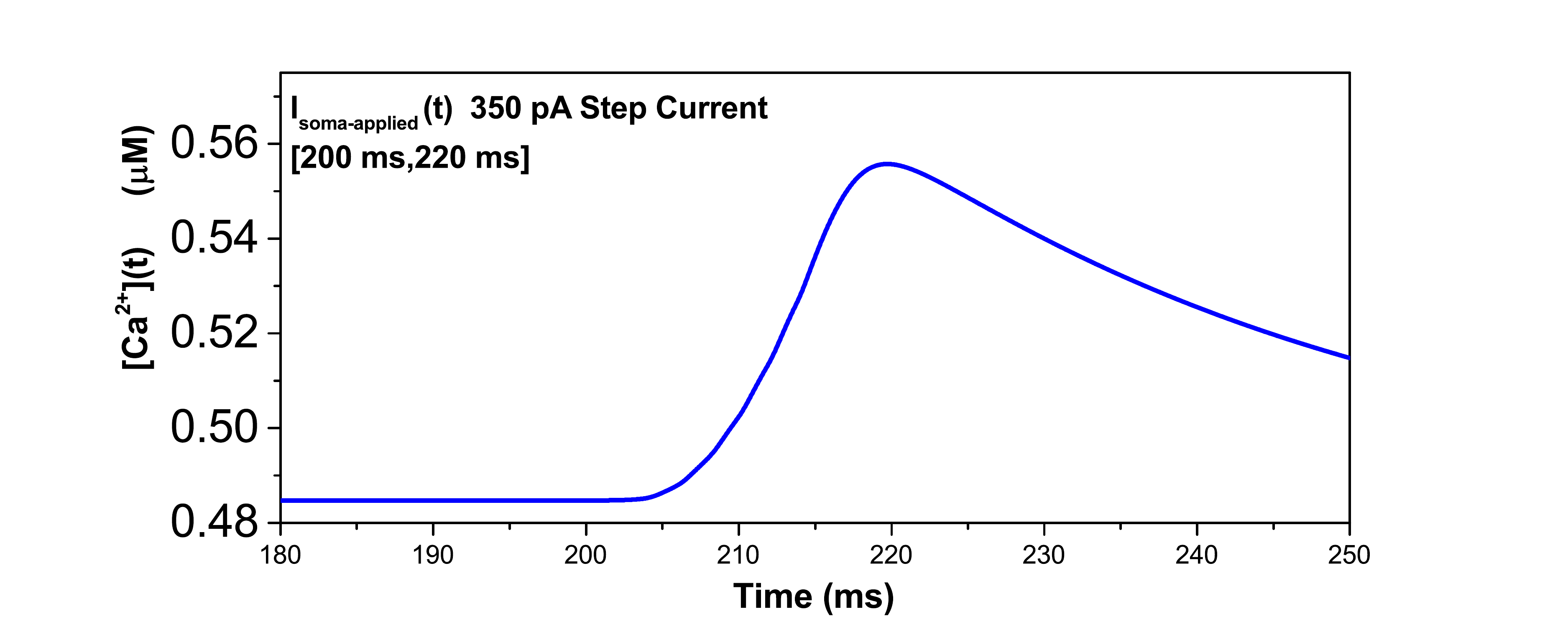} 
  \includegraphics[width=.49\textwidth]{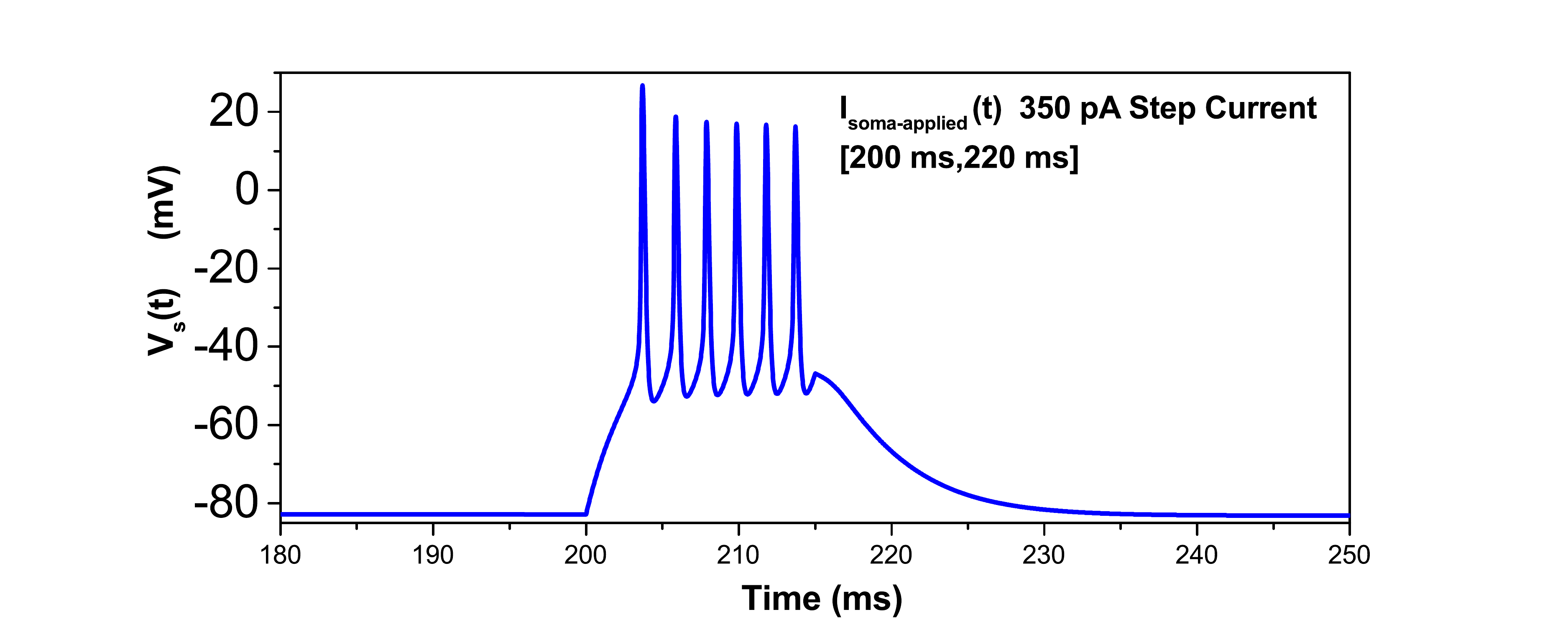} \\
  \caption{Increased injected current into the soma fails to elicit calcium spikes (left column). As the magnitude of a 20 ms step current pulse is increased from 150 pA (top graph) to 350 pA (bottom graph), the number of spikes increases steadily, rather than in an all-or-none fashion evoked by dendritic current stimulation}
  \label{fig:soma_bursting_1}
\end{figure*}

%%%% 			Figure 6 

\begin{figure*}
  \centering
  \includegraphics[width=.49\textwidth]{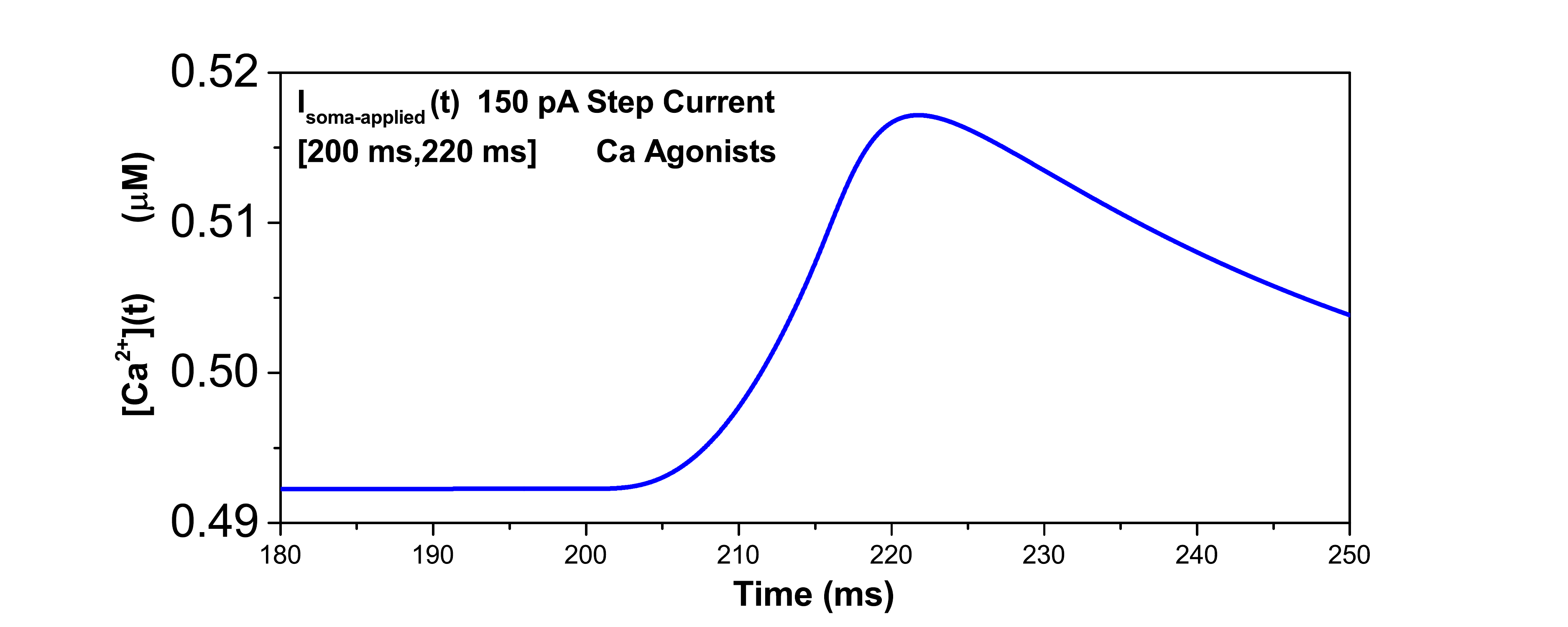} 
  \includegraphics[width=.49\textwidth]{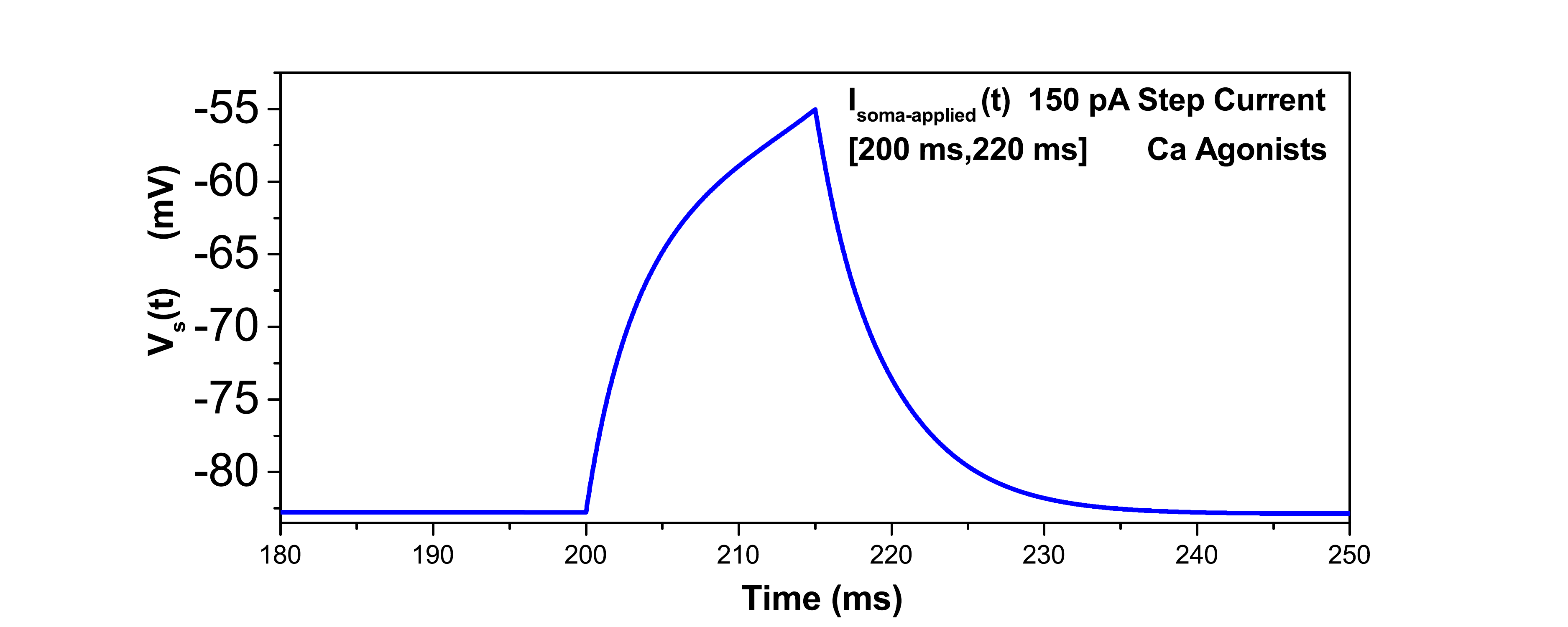} \\
  \includegraphics[width=.49\textwidth]{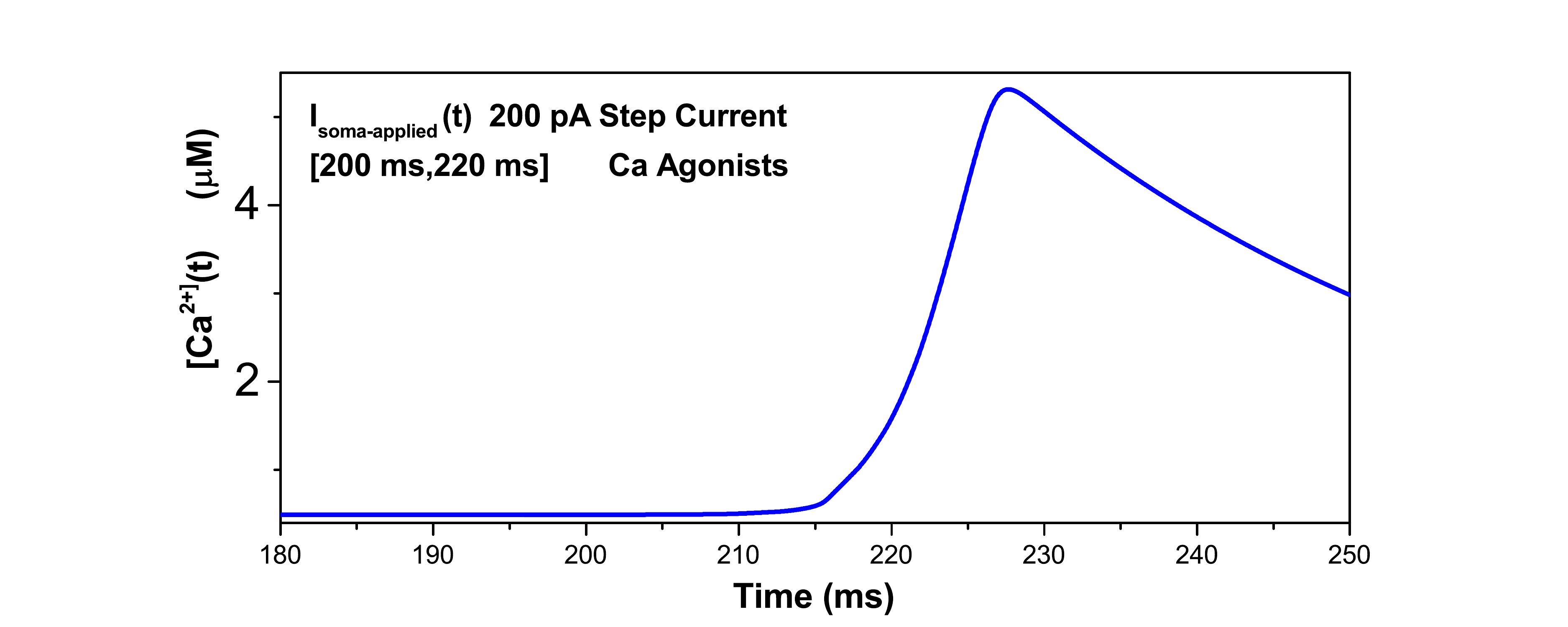} 
  \includegraphics[width=.49\textwidth]{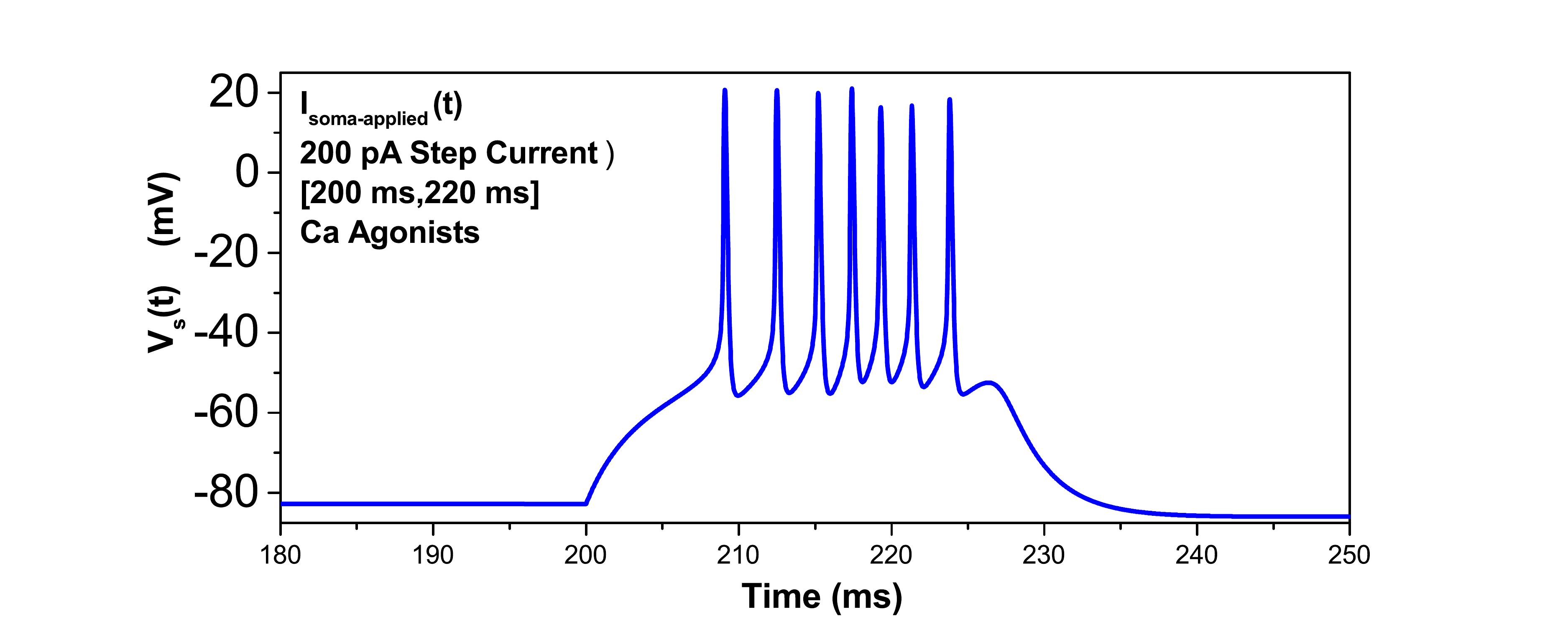} \\
  \includegraphics[width=.49\textwidth]{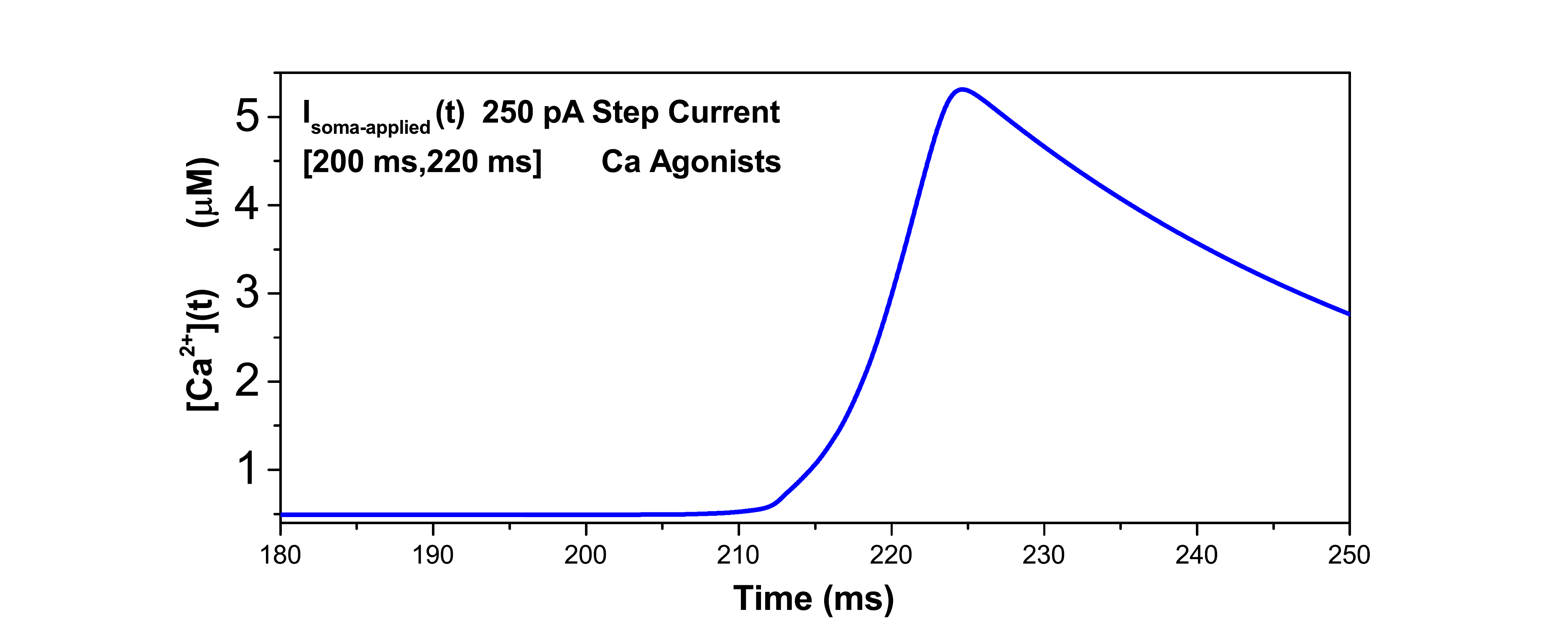} 
  \includegraphics[width=.49\textwidth]{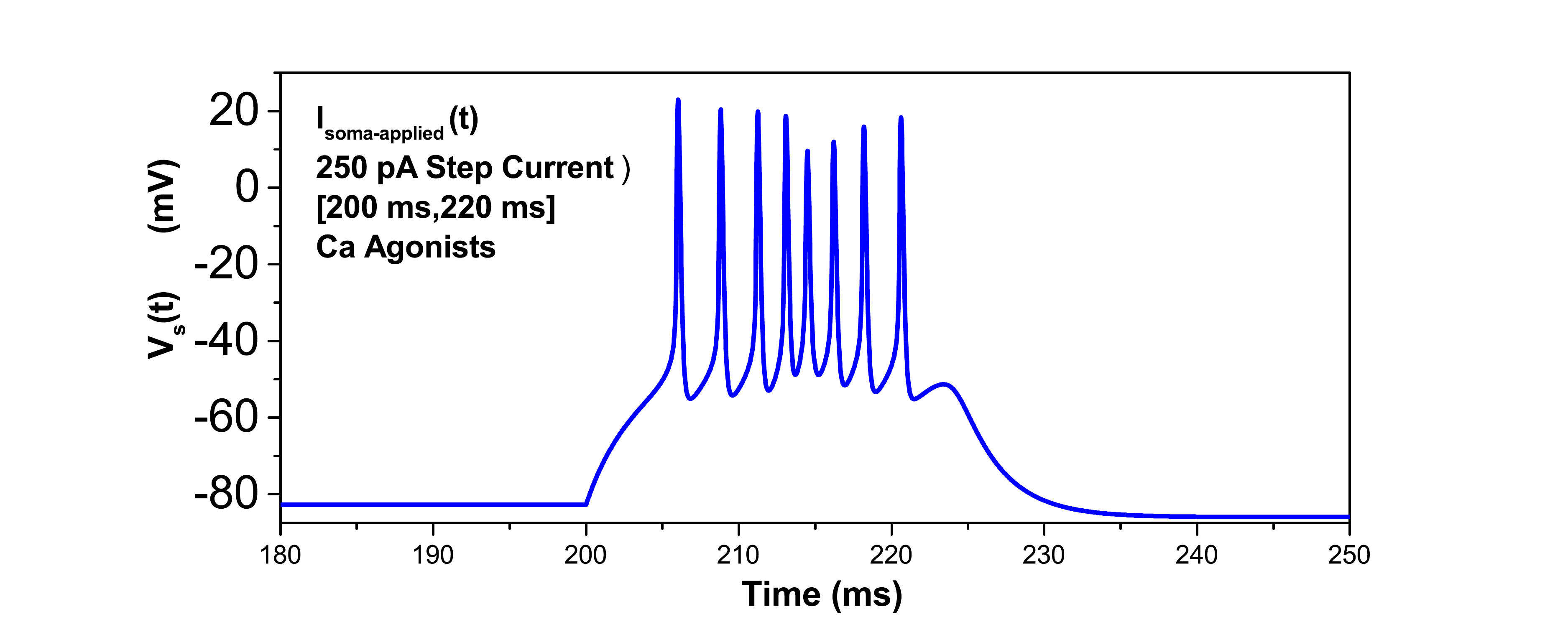} \\
  \includegraphics[width=.49\textwidth]{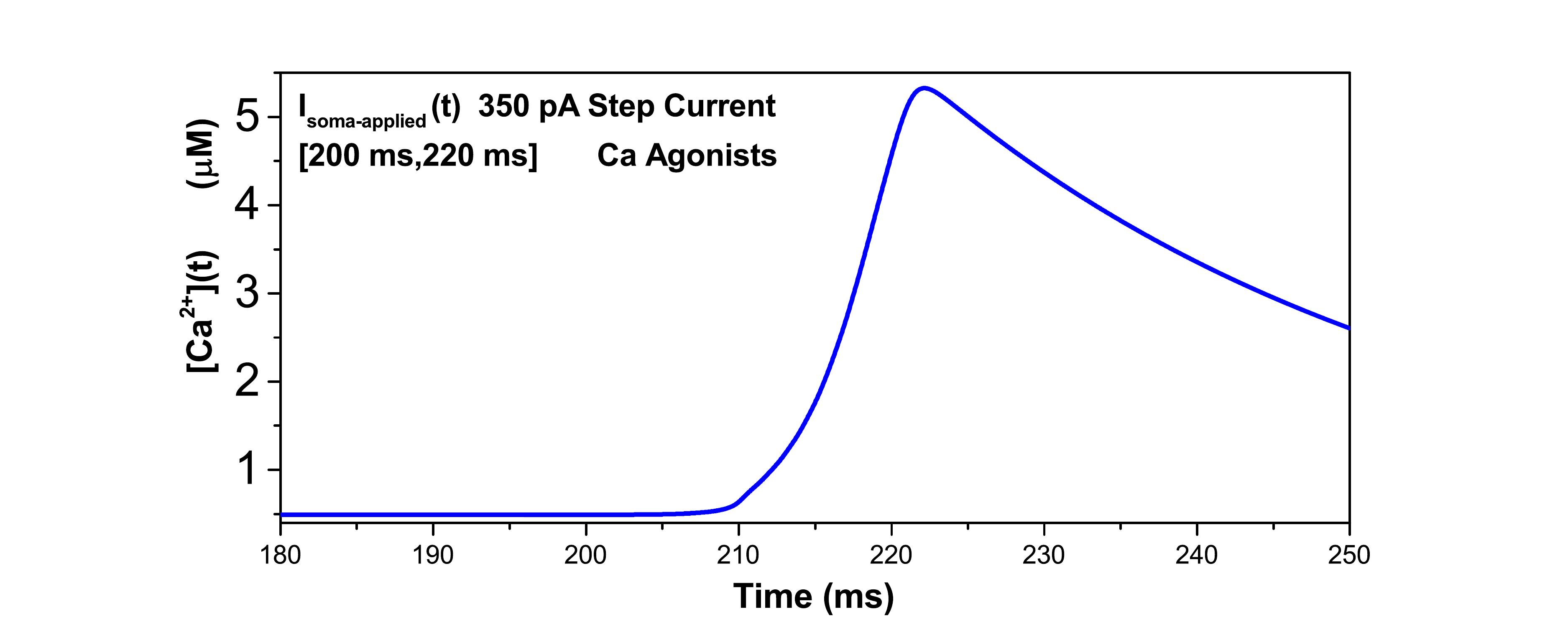} 
  \includegraphics[width=.49\textwidth]{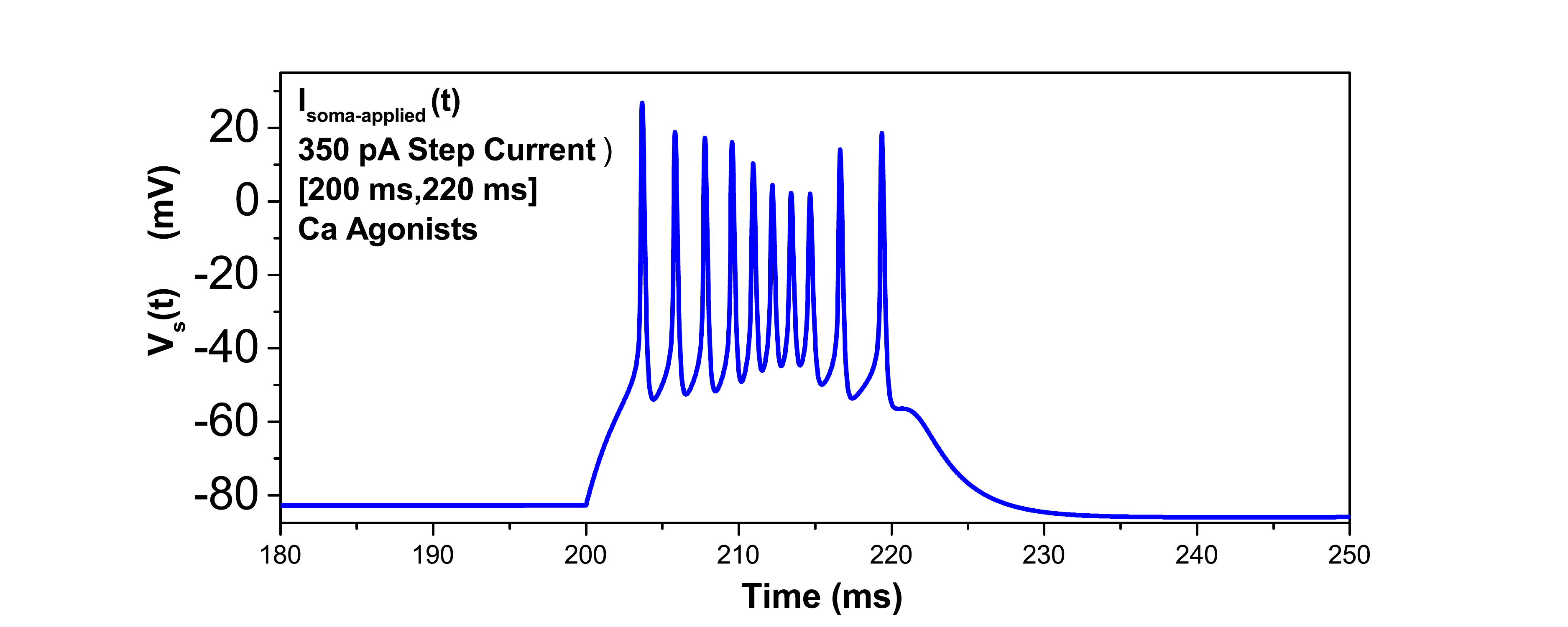} \\
  \caption{ In the presence of calcium agonists increased injected current into the soma can elicit bursting in the soma. As the magnitude of a 20 ms step current pulse is increased from 150 pA (top graph) to 350 pA (bottom graph), an all-or-none burst, coincident with a calcium spike, is invoked with sufficient current injection. This is in accord with experiment~\protect\cite{Long_2010}}
  \label{fig:soma_bursting_2}
\end{figure*}

It was further reported that current injection into the soma (as opposed to dendritic injections arising from synaptic currents) cannot produce all-or-none bursting \cite{Long_2010}, a feature we next investigated in our model. Step currents of increasing magnitude evoke progressively higher frequency somatic spiking, failing to elicit dendritic calcium spikes that lead to somatic bursts (Fig. \ref{fig:soma_bursting_1}). However, when the $g_{Ca-L}$ is increased -- simulating the presence of a calcium agonist -- dendritic calcium spikes and their accompanying all-or-none somatic bursts are evoked, again in line with experimental results \cite{Long_2010}. This is shown in Fig. \ref{fig:soma_bursting_2}.

Our model thus reproduces a large variety of qualitative features for this particular choice of parameters. We stress that this is no indication that these parameters are in any way correct for HVC\textsubscript{RA} neurons in the intact biological network or individual neurons in vitro. On the contrary, the main point is that the parameters will instead be determined by data assimilation. We only strive to demonstrate the plausibility of the model in that some chosen set of parameters qualitatively reproduce experimental findings. 

\subsection{Data Assimilation in a Reduced Model of HVC\textsubscript{RA} Neurons}
\label{sec: dendrite_only}

We have found that the annealing data assimilation methods developed above do not correctly estimate all of the parameters and unmeasured state variables in the HVC\textsubscript{RA} model when required to do so all at the same time. We attribute this to the difference in time scales of the fast spiking processes and the slower Ca dynamics. As such, we propose an alternate, feasible experimental protocol which allows the separation of these fast and slow processes. This is described in the following manner. 

One can envision the two-compartment model as containing fast spiking currents, K and Na, periodically modulated on a slower scale by the Ca-dependent K/Ca current. The time constant of intracellular Ca relaxation is likely between 15-50 ms, three to ten times the spiking time in the soma. If we decouple the neuron into fast and slow terms by setting the Na and K conductances to zero, effectively shutting off the fast currents, it may be possible in the nonlinear optimization routine to estimate the parameters governing the slow variation alone. Then, setting these parameters for the slow variation (the Ca dynamics) fixed in the full model, the remaining parameters governing the fast variables can be estimated through a second optimization. 

The data assimilation is thus carried out with a pair of incremental steps, each of reduced dimensionality. Experimentally, this two-step procedure can be realized by measuring the soma voltage trace of HVC\textsubscript{RA} neurons {\em in vitro}, in response to a user-defined somatically injected current, and then repeating the same measurement but now in the presence of Na and K channel blockers. These two data sets are then used in succession to determine first the slow parameters and then the fast ones.

To carry out the first step of this procedure, we generate data using the parameters listed in Table \ref{tab:qualitative_behavior}, with $g_{Na}$ and $g_K$ set to zero, using the injected somatic current shown in Fig. \ref{iappliedsoma}. Since $n(t)$, $m(t)$, and $h(t)$ are therefore decoupled from the system, they will not be considered, leaving one measured variable, $V_s(t)$, and three unmeasured state variables, $V_d(t)$, $r(t)$, and [Ca]$(t)$. We add the caveat that we must hold $C\textsubscript{0}$ fixed for the assimilation procedure to succeed. We stress that this does not imply that experimentally inaccessible information such as intracellular background [Ca] must be known. In fact, as will be illustrated below, even if $C\textsubscript{0}$ is held at some incorrect value, the optimization succeeds. As will be explained later, we attribute this to a partial degeneracy of the system, and in practice, will pose no negative ramifications for the data assimilation procedure.

\begin{figure}
  \centering
  \includegraphics[width=.52\textwidth]{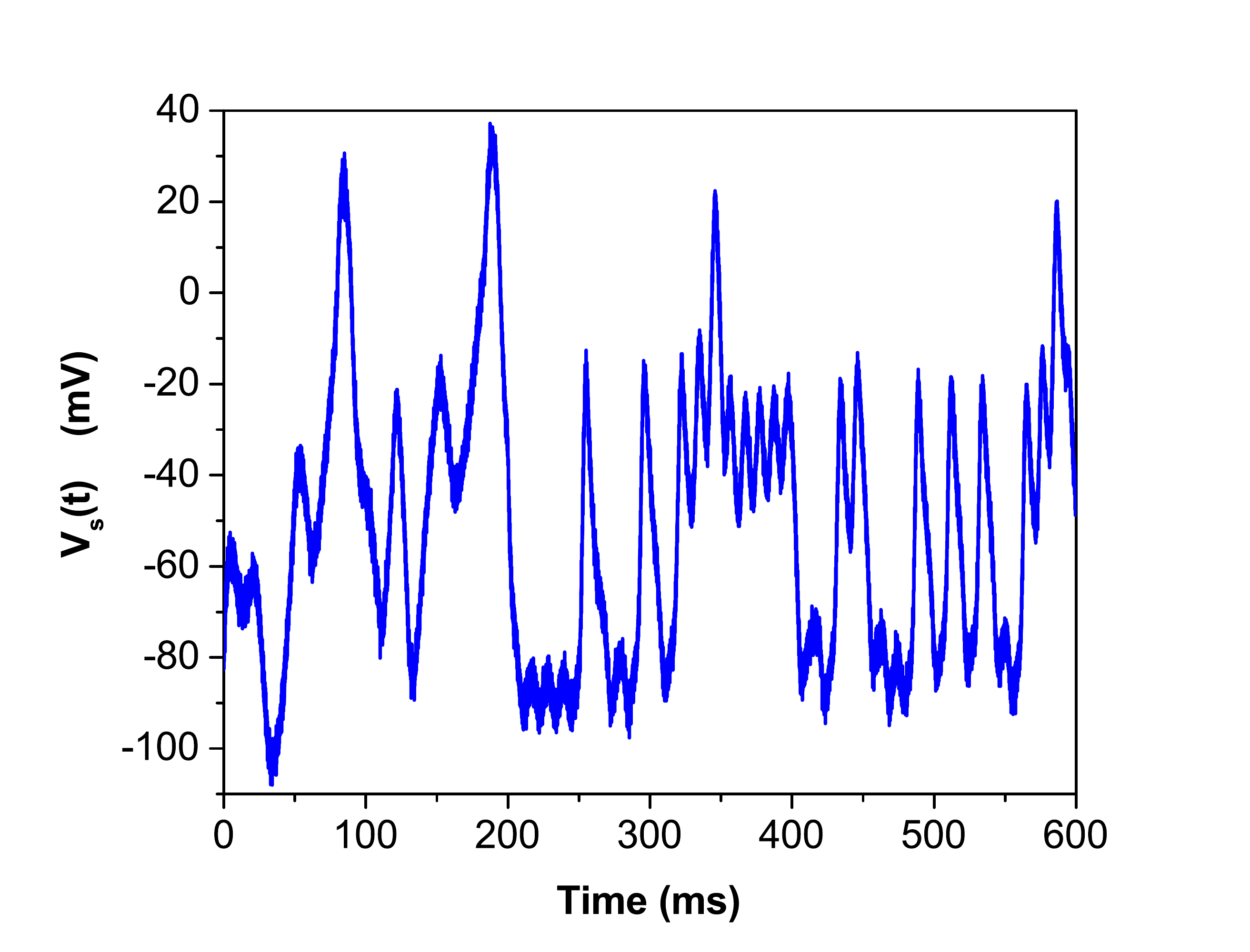} 
  \caption{The noisy data used in the data assimilation twin experiment}
  \label{fig:measured_data_dendrite_only}
\end{figure}

\begin{figure*}
  \centering
  \includegraphics[width=.95\textwidth]{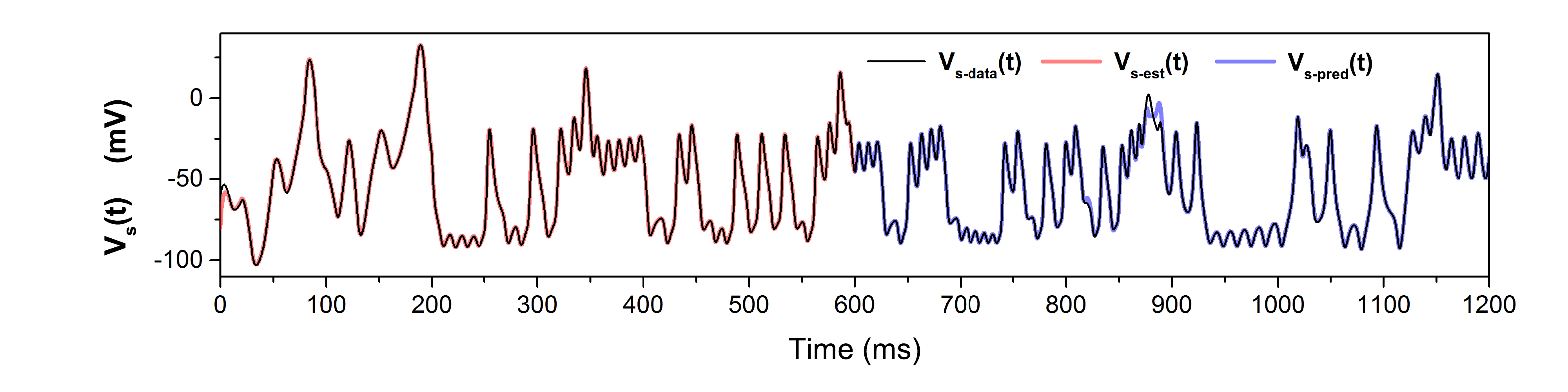} \\
  \includegraphics[width=.95\textwidth]{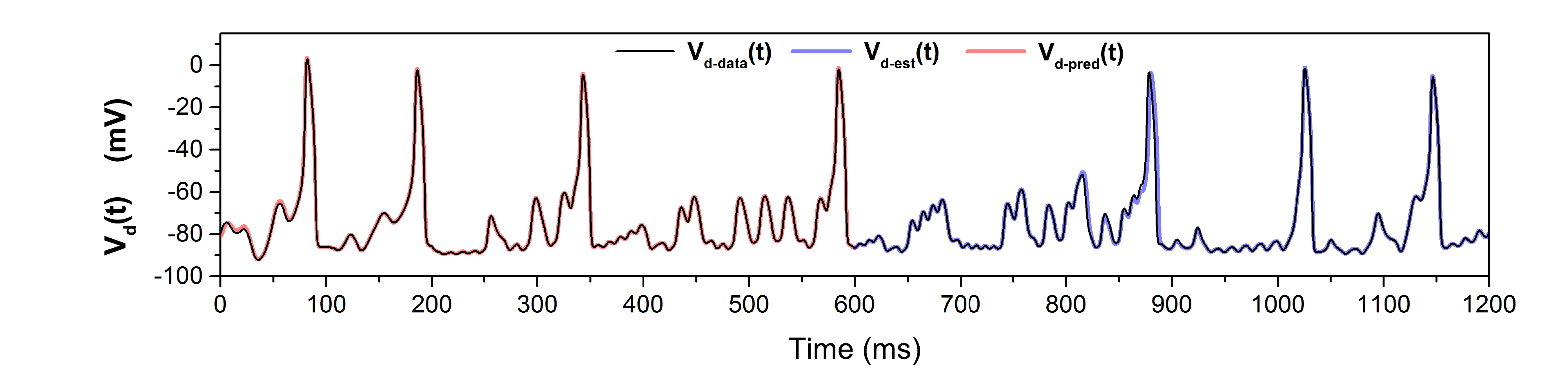}  \\   
  \includegraphics[width=.95\textwidth]{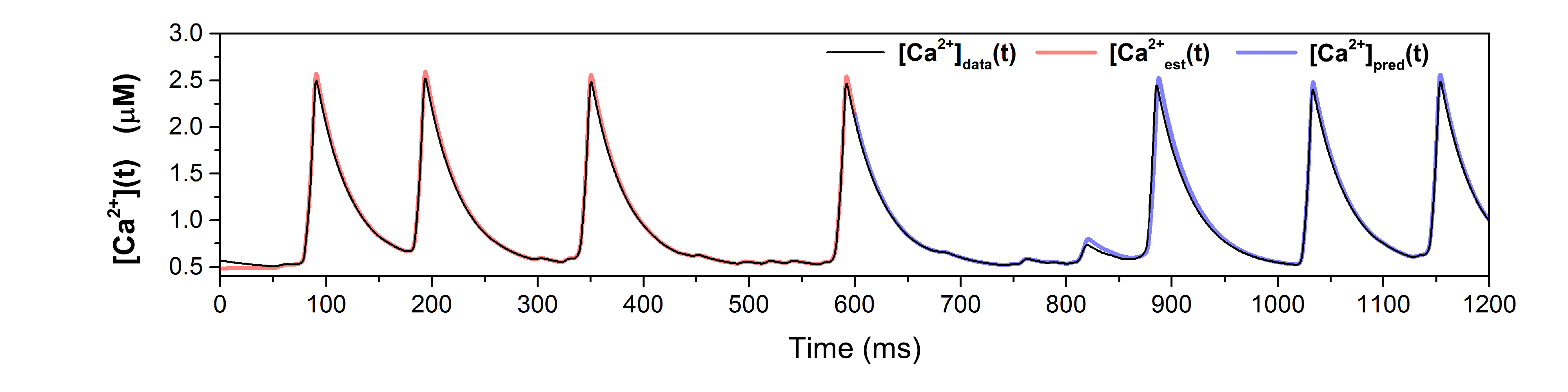}  \\
  \includegraphics[width=.95\textwidth]{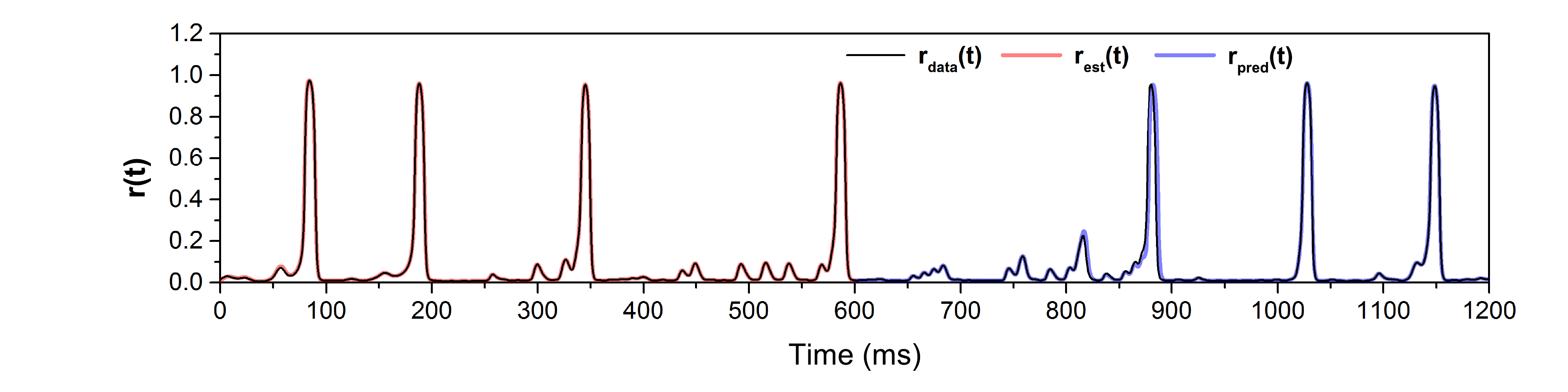} \\ 
\caption{ Estimation and prediction for the reduced model, with C\textsubscript{0} fixed at its correct value of 0.48. The estimated parameters are shown in Table \ref{tab:params_est_1}}
  \label{fig:dendrite_only_C0_fixed}
\end{figure*}

\begin{figure}
\centering
  \includegraphics[width=.48\textwidth]{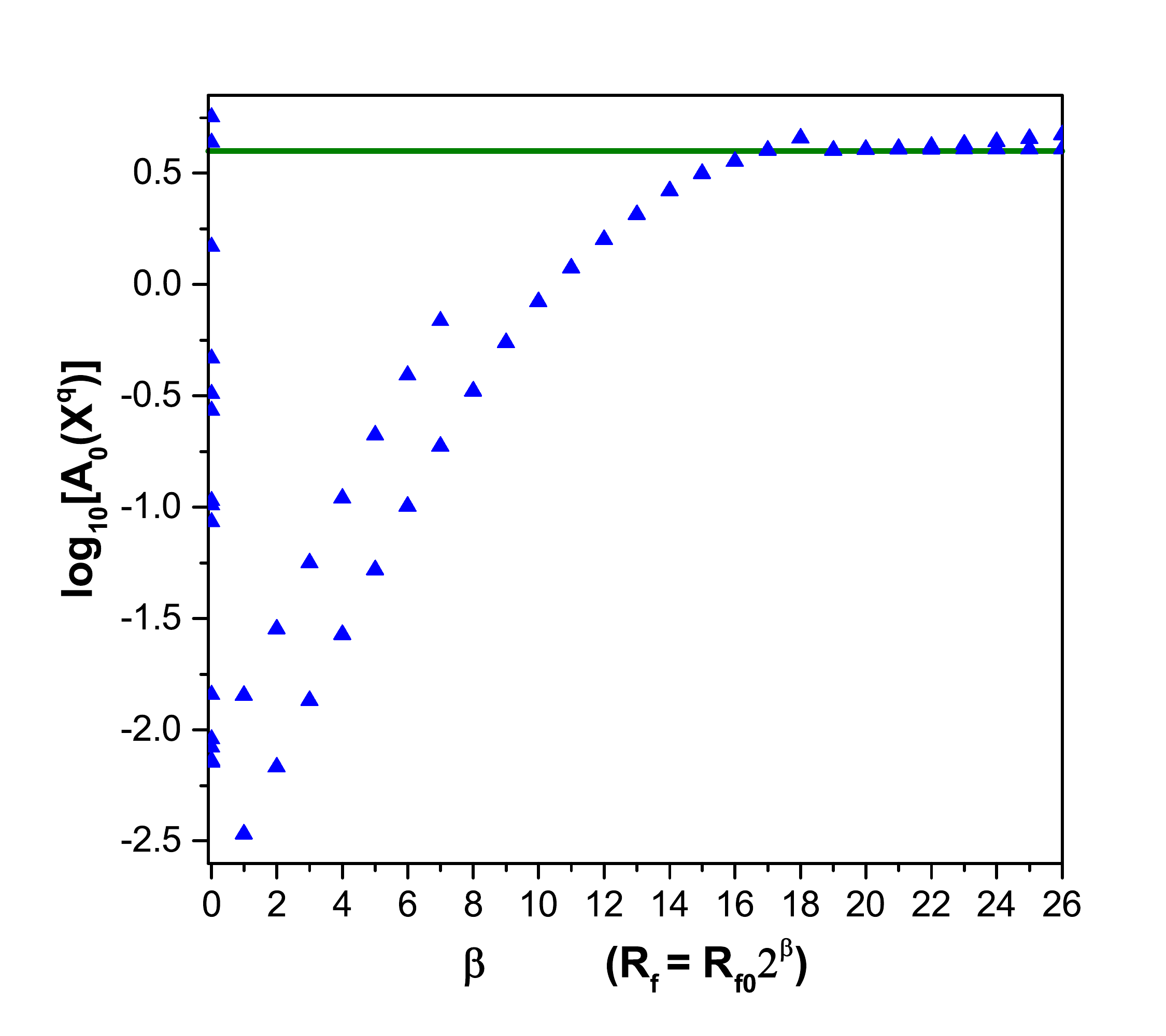} \\
  \includegraphics[width=.48\textwidth]{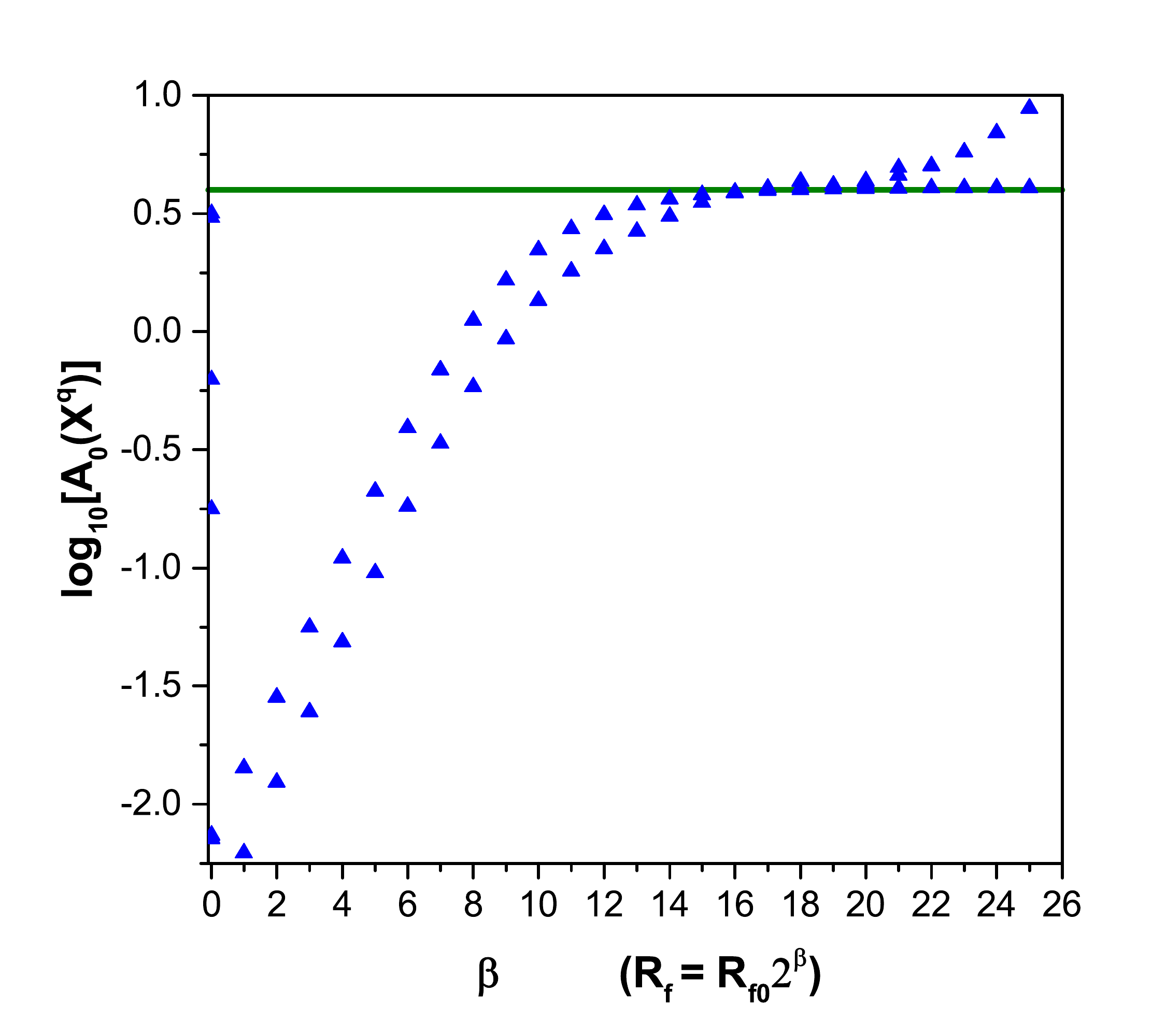}
  \caption{Action level plots for the dendrite only. {\it Top panel}. $C\textsubscript{0}$ fixed at its correct value. {\it Bottom panel}. Both $C\textsubscript{0}$ and [Ca]\textsubscript{ext} fixed at incorrect values. The horizontal line shows the action level of the lowest minimum of the action, which is achieved by a subset of the runs}
  \label{fig:dendrite_action}
\end{figure}

The generated data to be used in the data assimilation twin experiment (true voltage plus added Gaussian noise) is shown in Fig. \ref{fig:measured_data_dendrite_only}; the signal-to-noise ratio of the data is about 30 dB. The true path, estimation and prediction for the measured variable and three unmeasured variables are shown in Fig. \ref{fig:dendrite_only_C0_fixed}, and the corresponding action level plot is shown in Fig. \ref{fig:dendrite_action}. The true and estimated parameters, along with the search bounds for the nonlinear optimization, are listed in Table \ref{tab:params_est_1}. 

As is verified by the predictions of both unmeasured and measured state variables, the parameters are estimated to excellent accuracy. This is true despite the relatively lax bounds on the parameter search space, reflecting a relatively agnostic prior. We therefore expect the procedure to be robust to actual data, in which parameters such as conductances may indeed vary over several orders of magnitude.

%%%% 			Figure 7

\begin{table}
\centering
\begin{tabular}{c c c  c c  c}
\hline\noalign{\smallskip}
parameter & lower & upper & actual & estimated  \\
\noalign{\smallskip}\hline\noalign{\smallskip}
%& param. & lower & upper & actual & estimated\\
$E_{L}$& -110 & -70 & -80 & -80.08  \\
$E_K$  & -100& -75 &-90 & -89.91 \\
 $g_L$& 0.1 &10&3 & 3.03 	\\
$g_{Ca-L}$ & 0 & 10 & .06 & .1549\\
 $g_{Ca/K} $&  0 & 5000 & 240 & 250.35 \\
$g_{SD} $  & 1 & 50  & 5 & 4.96\\	
 $k_s$  &1 & 100  & 3.5 & 3.59\\
$ C_m$ & 1 & 100 & 21 & 21.02\\	
$\text{[Ca]}$\textsubscript{ext} & 1000 & 10000 & 2500 & 1000.0\\
$\phi $  &  1e-5 & 1e-2 & 8.7e-5 & 8.73e-5\\
$\theta_r$ &  -50 & -10 & -40 & -39.97 \\	
$\sigma_r$& 5 & 25 & 10 & 10.00 \\
$\theta_{\tau,r}$& -50 & -10 & 	--- & -50.00  \\
$\sigma_{\tau,r}$ & 5 & 25 &   --- & 5.00\\
$\tau_{r0}$  &0 & 1 & 1 & 1.000	\\
$ \tau_{Ca}$& 20 & 50 & 33 & 32.53\\
$\tau_{r1}$ &  0 & 1 & 0 & .0746 \\
$\tau_{r2}$ & 0 & 1 & 0 & 0.000	\\
\noalign{\smallskip}\hline
\end{tabular}
\caption{Estimated and true parameters, along with parameter bounds of the optimization, for the run with $C\textsubscript{0}$ fixed at 0.48 and the Na and K channels shut off. The parameters $\theta_{\tau,r}$ and $\sigma_{\tau,r}$ are not relevant since the coefficient for the terms that involve them ($\tau_{r1}$ and $\tau_{r2}$) are zero for the generated data; the data is insensitive to their value. For this reason their ``actual values" are blank, and it is not surprising that their estimated values lie at the parameter bounds. The parameter units are as follows: mV for all $\theta_i$, $\sigma_i$, and $E_i$; ms for all $\tau_i$; nS for $g_L$, $g_{Ca/K}$, and $g_{SD}$; $\mu$M/pA/ms for $g_{Ca-L}$; pF for $C_m$; $\mu$M for $k_s$ and [Ca]\textsubscript{ext}}
\label{tab:params_est_1}
\end{table}

%%%			Figure 8

%%%			Figure 9

We note that a few of the parameters were estimated incorrectly, despite the accurate predictions. This indicates that these parameters are degenerate, that is, different values may result in similar predicted trajectories. In particular, $g_{Ca-L}$ and Ca\textsubscript{ext} are far from their actual values, but their product, which appears as the dominant term in the calcium current, is still the same ($2500 \times .06 = 1000 \times .15$). The same situation appears with $k_s$ and $g_{Ca/K}$, which appear only in the products $\frac{k_s^2}{k_s^2 + [\text{Ca}](t)^2}g_{Ca/K}$. Thus, compensating shifts in these parameters can produce the same behavior to high precision. This implies that it may be difficult or impossible to determine the individual parameters exactly in practice. 

\begin{figure*}
  \centering
  \includegraphics[width=.95\textwidth]{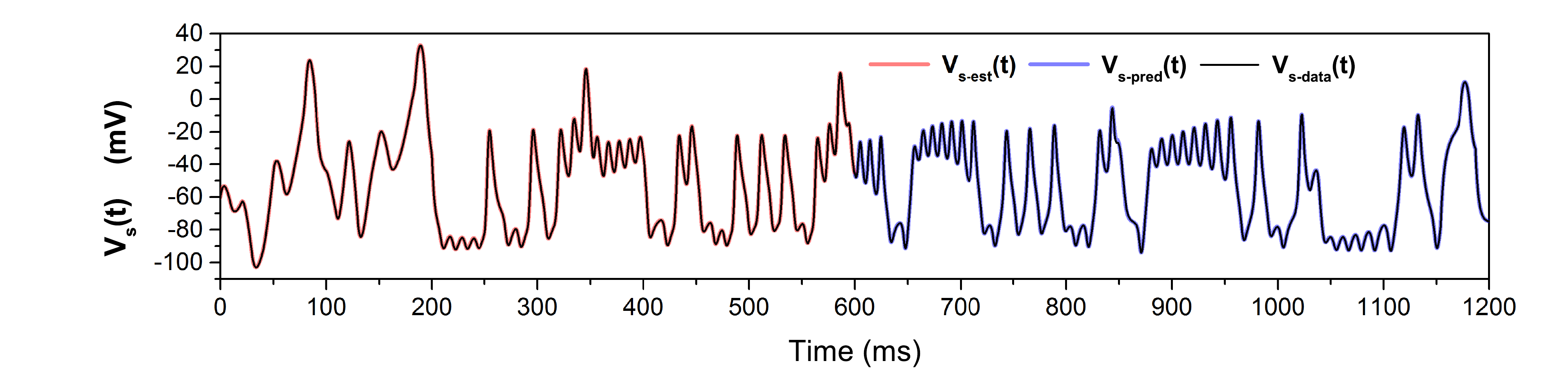} \\
  \includegraphics[width=.95\textwidth]{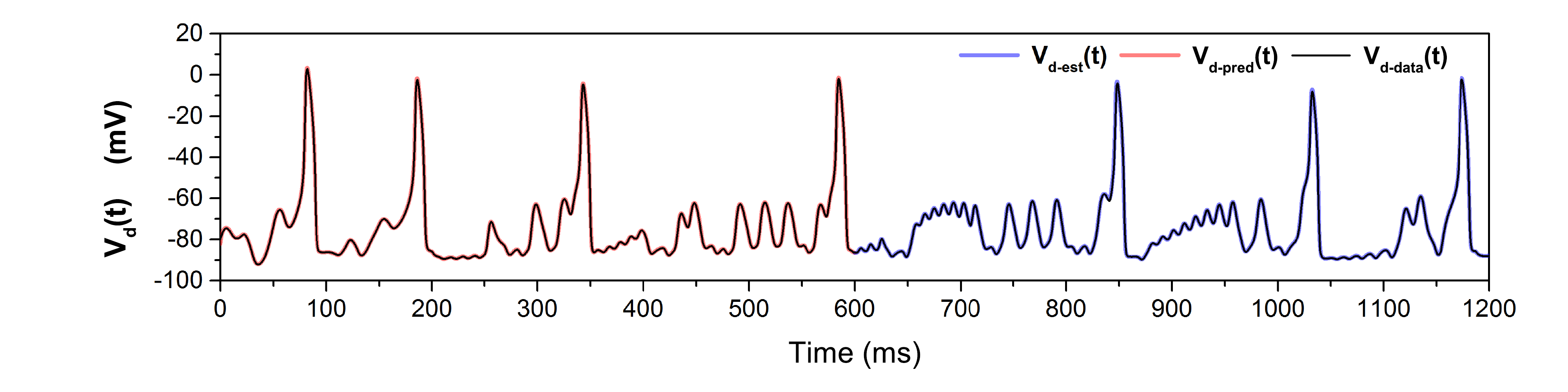}  \\   
  \includegraphics[width=.95\textwidth]{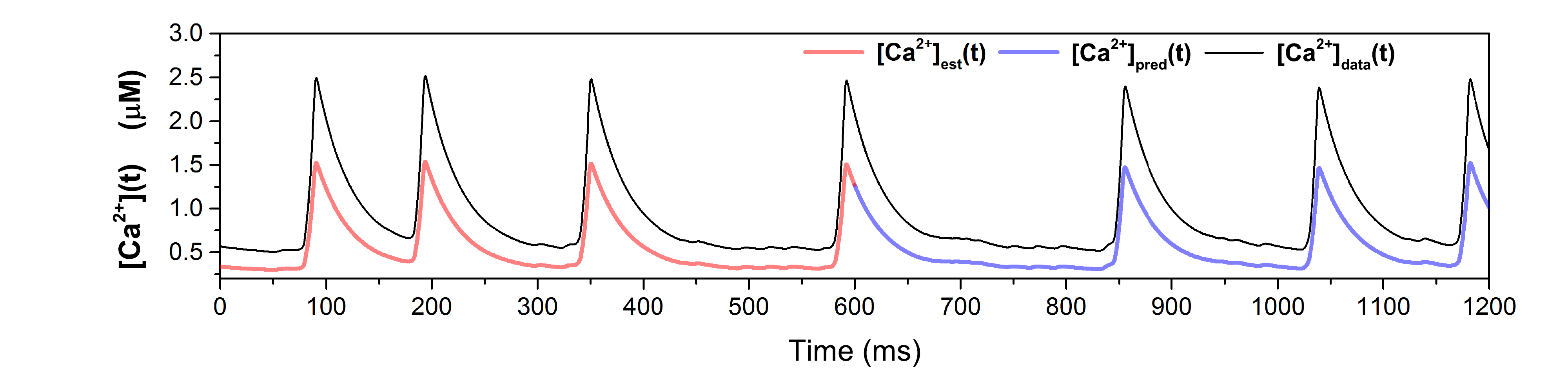} \\ 
  \includegraphics[width=.95\textwidth]{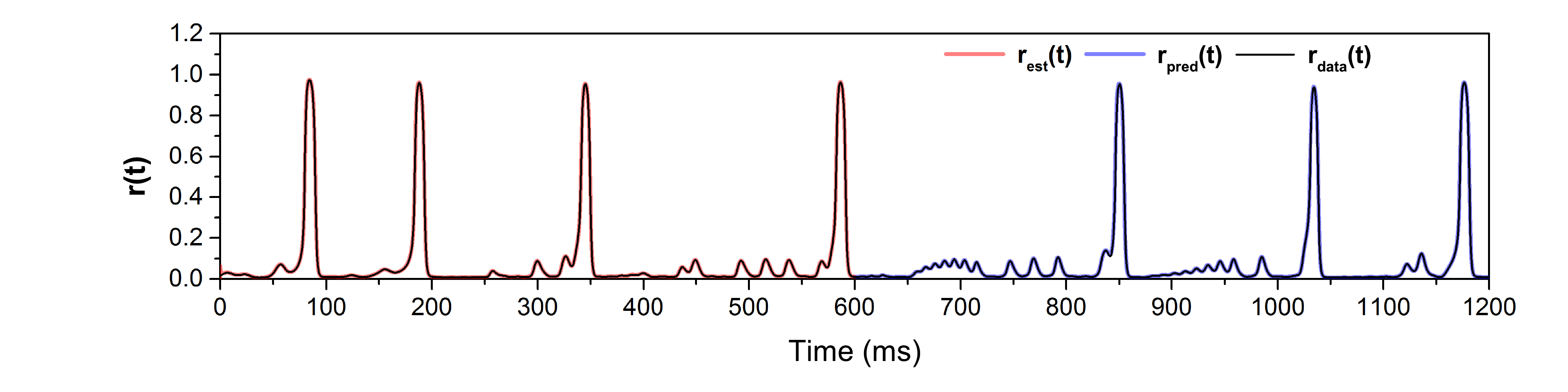}  \\
\caption{Estimation and prediction for the reduced model, with parameters listed in Table \ref{tab:params_est_2}. Two parameters are fixed at incorrect values but the same order of magnitude as the correct values. The predictions are excellent for all variables, except [Ca]$(t)$, which is qualitatively correct but shifted lower, due to the erroneous background concentration. Despite this, the predictions for the other variables (voltages and gating variables) are excellent}
  \label{fig:dendrite_only_wrong_params}
\end{figure*}

The requirement that $C_{0}$ be fixed to its correct value may be relaxed. For example, fixing both $C\textsubscript{0}$ and Ca\textsubscript{ext} at the incorrect values of $0.285$ and $1000$ $\mu$M, respectively, again results in excellent predictions, as seen in Fig. \ref{fig:dendrite_only_wrong_params}. Only [Ca]$(t)$ results a qualitatively similar but shifted trace, due to the lowered background concentration. The estimated parameter values are excellent, with discrepancies only in $g_{Ca-L}$ and $k_s$ for the reasons indicated above, and in a reduced $\phi$ value to balance the smaller [Ca] decay term $\frac{C\textsubscript{0}}{\tau_{Ca}}$ in the [Ca] dynamics (Table \ref{tab:params_est_2}). As in the previous case, many, but not all of the initial runs achieve the global minimum. This is shown in Fig. \ref{fig:dendrite_action}.

%%%%			Figure 10

%%% 			Table 3

\begin{table}
\centering
\begin{tabular}{ c  c  c  c  c }
\hline\noalign{\smallskip}
parameter & lower& upper & actual & estimated \\
\noalign{\smallskip}\hline\noalign{\smallskip}
 $g_L$  & 0.1 &10&3 & 3.03  		 \\
$g_{Ca-L}$ & 0 & 10 & .06 & .155\\
$E_{L}$ & -110 & -70 & -80 & -80.10  		 \\
$k_s$&1 & 100  & 3.5 & 2.10\\
$E_K$ & -100& -75 &-90 & -89.91	 	\\
$g_{SD} $ & 1 & 50  & 5 & 4.96\\	
$\theta_r$& -50 & -10 & -40 & -39.90	 	\\
$ C_m$ & 1 & 100 & 21 & 21.01\\	
$\sigma_r$& 5 & 25 & 10 & 10.00  	 	\\
$ \tau_{Ca}$ & 20 & 50 & 33 & 33.13\\
$\theta_{\tau,r}$& -50 & -10 & 	--- & -50.00	 	\\
$g_{Ca/K} $&0 & 5000 & 240 & 242.51 \\
$\sigma_{\tau,r}$& 5 & 25 &   --- & 5.00			\\
$\phi $ & 1e-5 & 1e-2 & 8.7e-5 & 5.22e-5\\
$\tau_{r0}$ &0 & 1 & 1 & .99			\\
$\tau_{r2}$ &  0 & 1 & 0 & 0.000	\\		
$\tau_{r1}$  & 0 & 1 & 0 & 0.08 		\\
\noalign{\smallskip}\hline
\end{tabular}
\caption{Estimated and true parameters, along with the search bounds in the optimization procedure, for the reduced model. Here, $C\textsubscript{0}$ and Ca\textsubscript{ext} were held fixed at the incorrect values of 0.285 and 1000 $\mu$M, respectively. The units are the same as those of Table \ref{tab:params_est_1}}
\label{tab:params_est_2}
\end{table}

\subsection{Data Assimilation in an HVC\textsubscript{RA} Neuron, Using Estimated Dendritic Parameters}
\label{sec: soma_fixed_dendrite}

\begin{figure}
\centering
  \includegraphics[width=.48\textwidth]{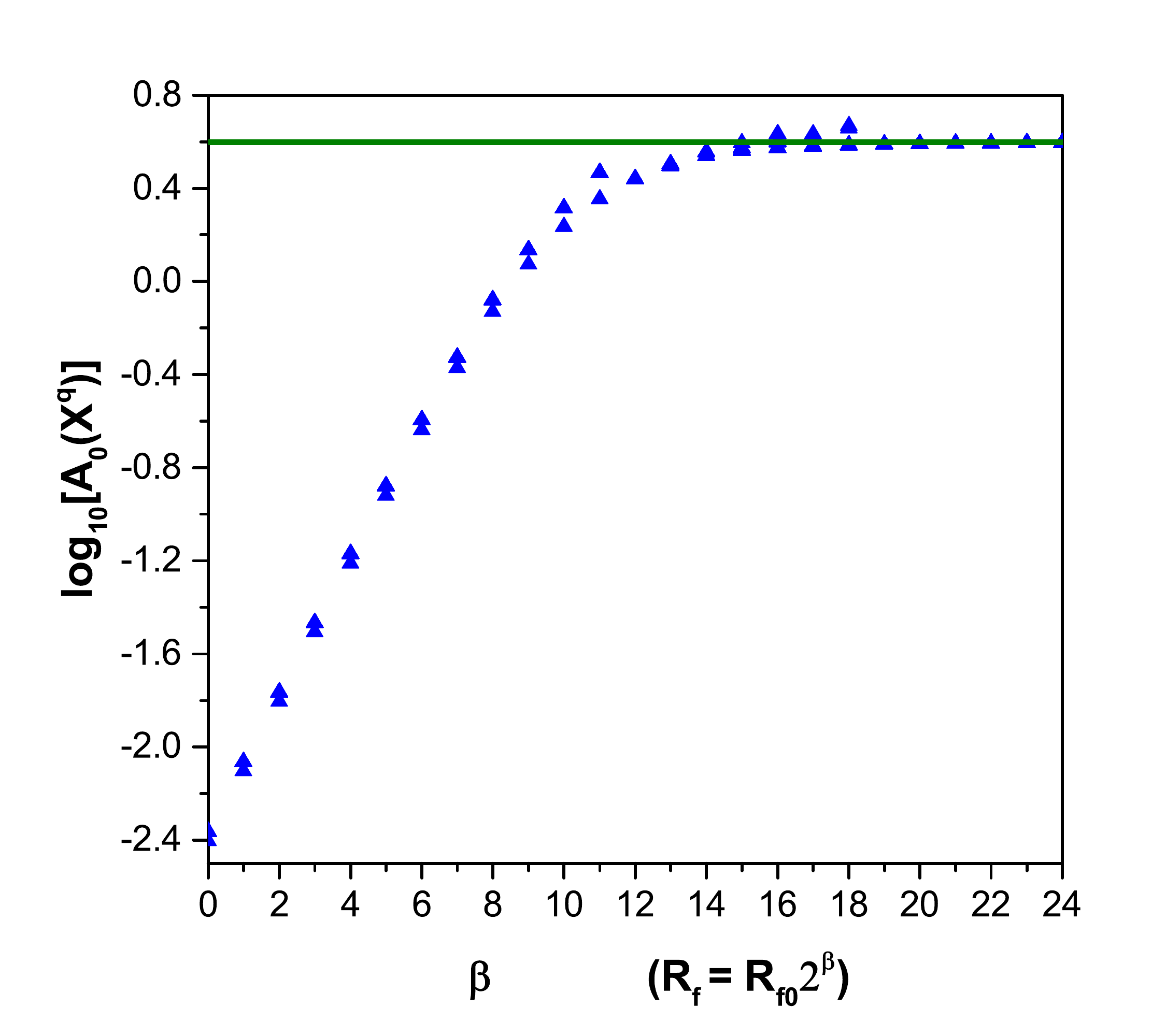}
  \caption{Action levels from the annealing process when dendritic parameters are fixed to the values listed in Table \ref{tab:params_est_2}, and the remaining parameters of the full model are estimated}
  \label{fig:soma_action}
\end{figure}

\begin{figure*}
  \centering
  \includegraphics[width=.95\textwidth]{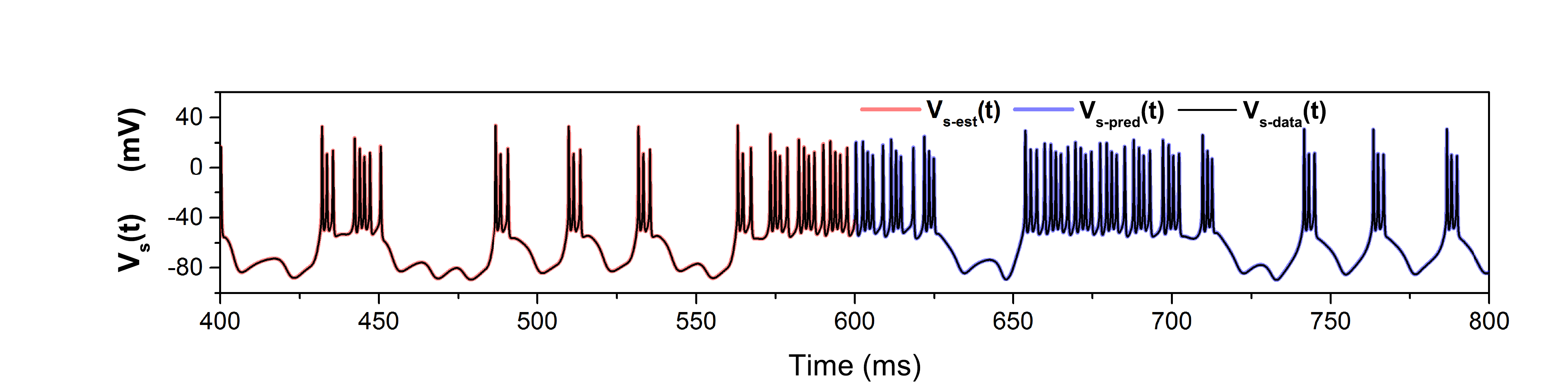} \\
  \includegraphics[width=.95\textwidth]{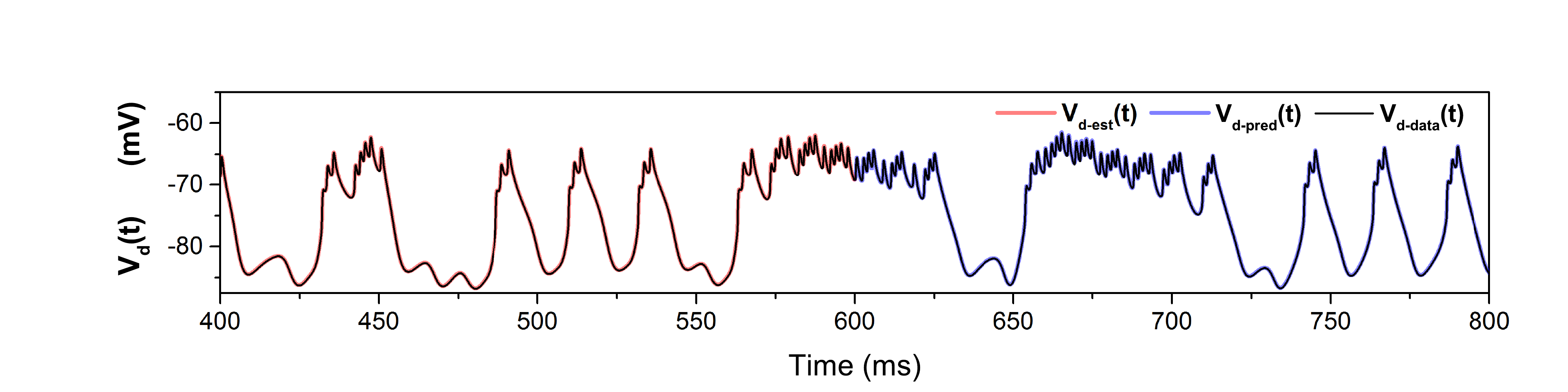} \\
  \includegraphics[width=.95\textwidth]{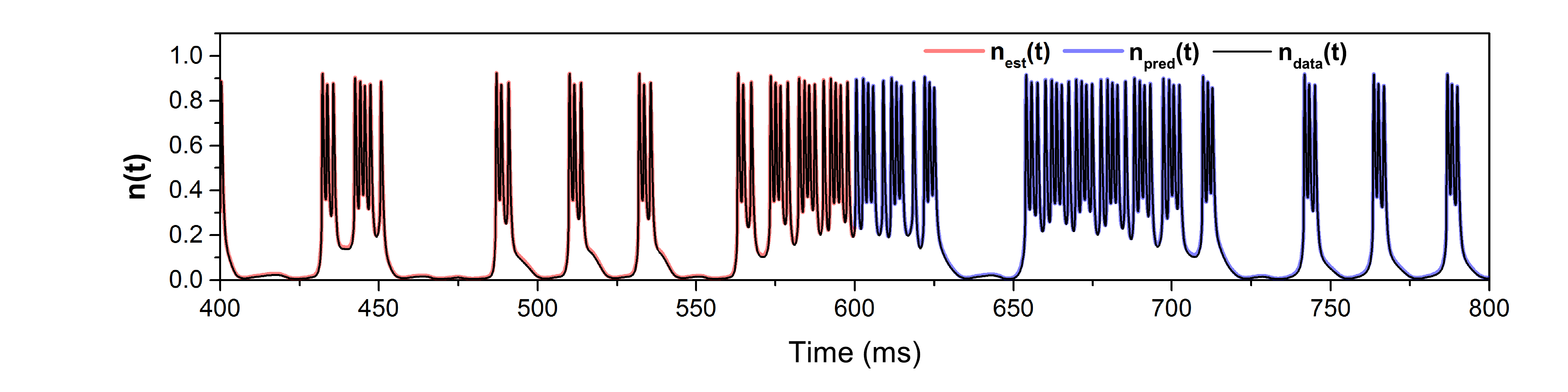} \\
  \includegraphics[width=.95\textwidth]{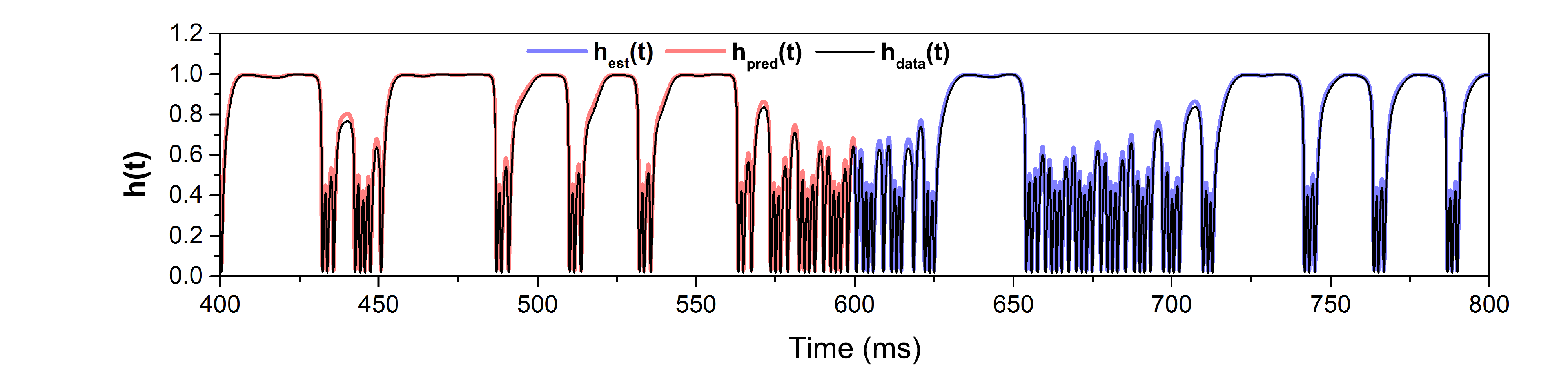} \\
  \includegraphics[width=.95\textwidth]{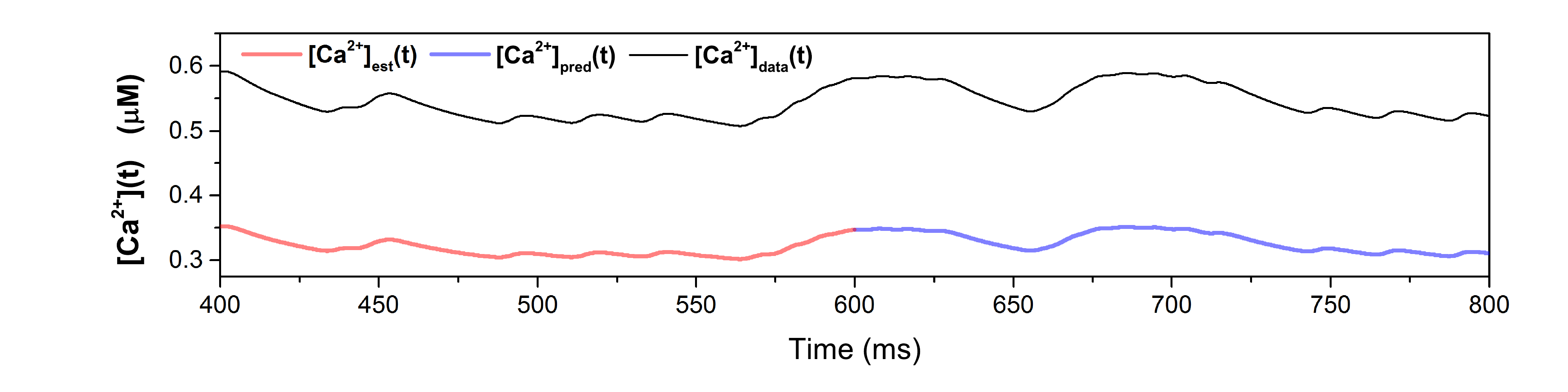} \\
\caption{Estimation and prediction for the full model, with estimated parameters listed in Table \ref{tab:params_est_3}, and the remaining parameters (the fast variables in the soma) fixed according to the estimations given in Table \ref{tab:params_est_2}. The predictions are excellent for the five state variables shown. Estimations and predictions for states not shown ($r(t)$ and $m(t)$) are similarly accurate}
  \label{fig:soma_fixed_dendrite}
\end{figure*}

We now illustrate how the estimates gathered from the reduced model with no Na or K channels can be used in determining the parameters and states of the full HVC\textsubscript{RA} neuron. The parameter estimates from the reduced model are now set fixed in the dynamical equations of the full model. For more generality, we used the values in Table \ref{tab:params_est_2}, with the erroneous values of $C\textsubscript{0}$ and Ca\textsubscript{ext}. The parameters $g_{Na}$ and $g_K$ are no longer held at zero, and there are six unmeasured state variables and one measured state variable. Action level plots for 50 runs are shown in Fig. \ref{fig:soma_action}, indicating that most of the runs have achieved the lowest minimum. Parameter estimates corresponding to runs that found the lowest minimum are listed in Table \ref{tab:params_est_3}, and the corresponding estimations and predictions are shown in Fig. \ref{fig:soma_fixed_dendrite}. Most parameters are estimated to high accuracy and the predictions are excellent for all state variables.

%%%%			Figure 11

%%%% 			Figure 12

%%%% 			Table 4		

\begin{table}
\centering
\begin{tabular}{c c c c c }
\hline\noalign{\smallskip}
parameter & lower& upper & actual & estimated \\
\noalign{\smallskip}\hline\noalign{\smallskip}
 $g_{K}$   &  0 &  5000  &   120   &    123.36 \\
$g_{Na}$ &   0   &  5000   &    1050    &   757.81 \\
$E_{Na}$  &   50   &   60   &    55    &   60.00 \\
$\theta_m$ &    -50   &  -10   &    -30    &   -30.44 \\
$\sigma_m$   &  6.25   &  16.67   &    9.5    &   9.68  \\
$\theta_{\tau,m}$ &   -40   &  -20   &    ---    &  -29.92\\
 $\sigma_{\tau,m}$   &   -50   &  -5   &    --- & -8.24 \\
 $\tau_{m0}$  &   0  & 1  &   .01  &    7.23e-4   \\
 $\tau_{m1}$  &0  & 1  &  0 &   5.46e-3   \\
 $\tau_{m2}$  & 0  & 1  &  0   &    0.000    \\
$\theta_n$  & -50   &  -10   &    -35    &   -34.28 \\
$\sigma_n$  & 6.25   &  16.67   &    10    &   10.82 \\
$\theta_{\tau,n}$  &   -40   &  -20   &    -27    &  -32.41 \\
$\sigma_{\tau,n}$  & -50   &  -5   &    -15    &  -12.41 \\
$\tau_{n0}$  &  0  & 1  &    .1    &  .104 \\
$\tau_{n1}$  &  0  & 1  &    0   &  .100 \\
$\tau_{n2}$  &  0  & 1  &    .5    &  .42 \\
$\theta_h$  &  -50   &  -10   &    -45 &  -43.43   \\
$\sigma_h$  &  -16.67   &  -6.25  &    -7    & -7.05  \\
$\theta_{\tau,h}$  &  -50   &  -20   &   -40.5    & -41.30 \\
$\sigma_{\tau,h}$   & -50   &  -5   &    -6   & -5.00 \\
$\tau_{h0}$   &  0  & 1  &  .1 &  .12\\
$\tau_{h1}$  &   0  & 1  &    0  &   .04 \\
$\tau_{h2}$  &  0  & 1  &    0.75   & .69 \\
\noalign{\smallskip}\hline
\end{tabular}
\caption{Estimated and true parameters, along with the search bounds in the optimization procedure, in the full model. The parameters not listed here were fixed to the estimated values in the first assimilation step (Table \ref{tab:params_est_2}), along with the C\textsubscript{0} and  Ca\textsubscript{ext} held fixed at the incorrect values of 0.285 and 1000. $\mu$M, respectively. The units for all $\tau_i$ are ms; all other parameters have units of mV}
\label{tab:params_est_3}
\end{table}

%%%%%%%%%% 				Discussion 				%%%%%%%%%%

\section{Discussion}

In developing the model of the HVC\textsubscript{RA} neuron, we strove for the preservation of some features such as dendritically stimulated bursting and burst excitability dependence upon Ca channel strength, while other features such as hyperpolarization sag and spike rate adaptation are neglected. When carrying out this procedure with actual HVC\textsubscript{RA} data, inspection of the action level plots can expose these model deficiencies. Missing or incorrect current terms in the model are indicated by action levels that rise precipitously with increasing $\beta$ for large $\beta$ values, rather than converging on a limiting value. This mark of inconsistency is one of the advantages of the annealing method in the stationary path approximation to the path integral.

It has been shown that a combination of pharmalogical manipulation and data assimilation can be used to determine a large number of parameters and unmeasurable state variables in HVC\textsubscript{RA} neurons, given somatic voltage measurements alone. One could certainly propose a similar experiment in which, instead, the L-type Ca channels are blocked and the fast, soma parameters are determined via data assimilation -- these estimated parameters are then held fixed in the assimilation of the full model. We have found similarly excellent estimations and predictions for this twin experiment, not shown here, indicating that the success of this joint procedure is not specific to the particular combination of chemical manipulation and assimilation illustrated above.

%%%%			 Figure 13

\begin{figure}
  \centering
  \includegraphics[width=.48\textwidth]{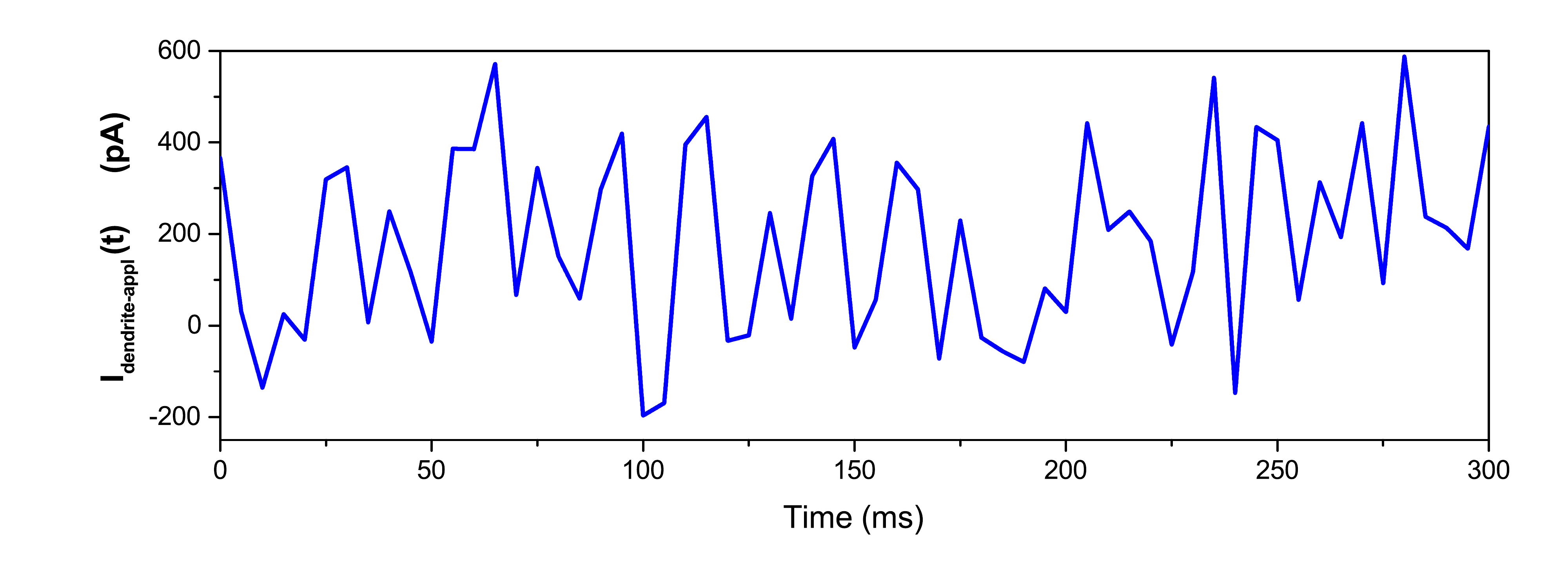} \\
  \includegraphics[width=.48\textwidth]{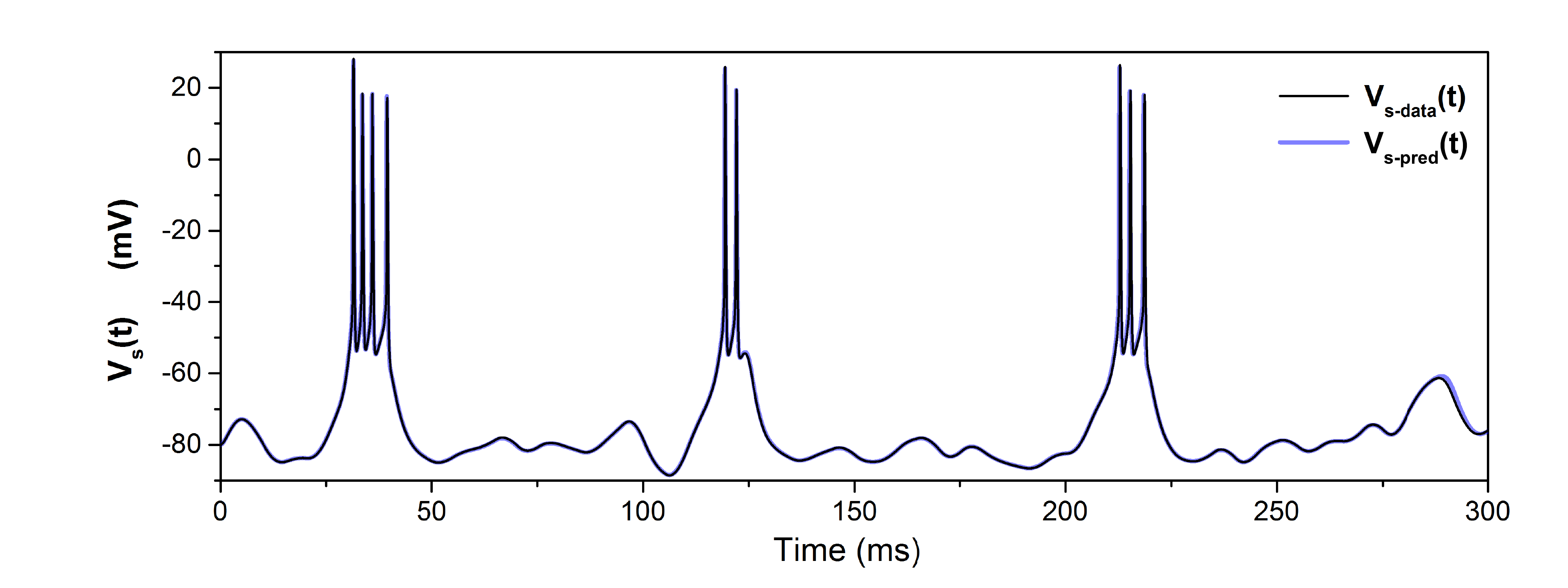} \\
  \includegraphics[width=.48\textwidth]{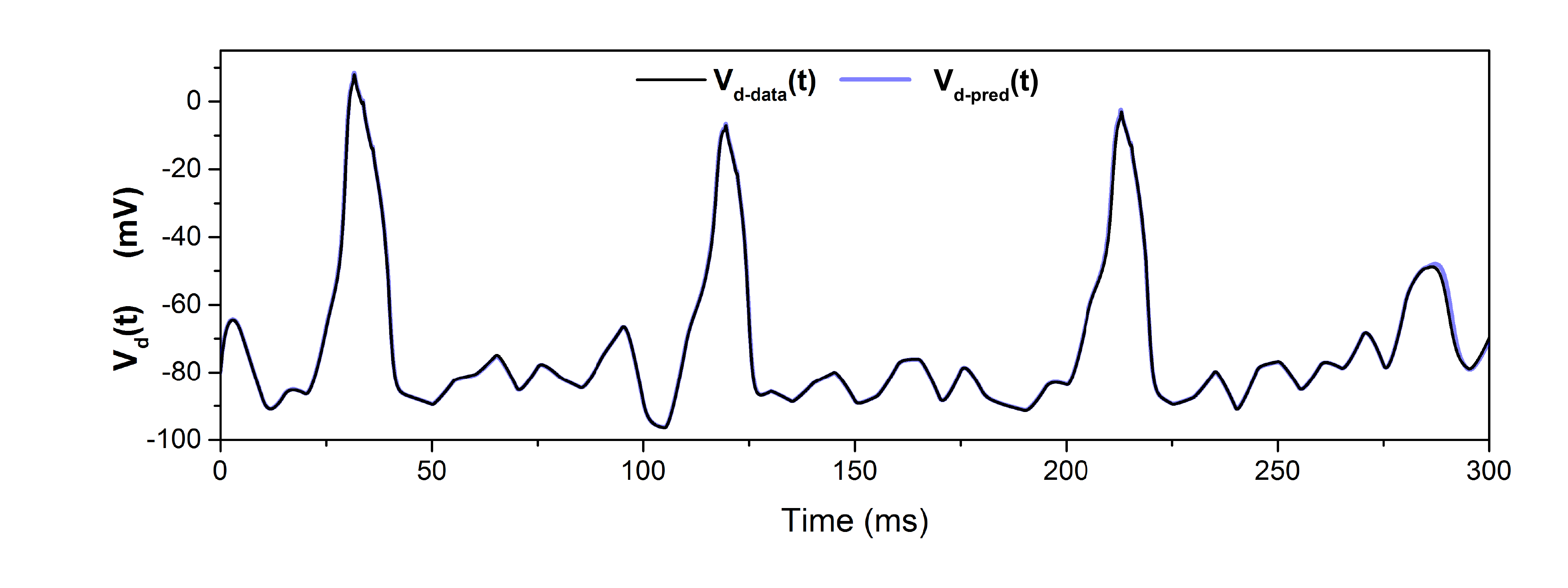} \\
  \includegraphics[width=.48\textwidth]{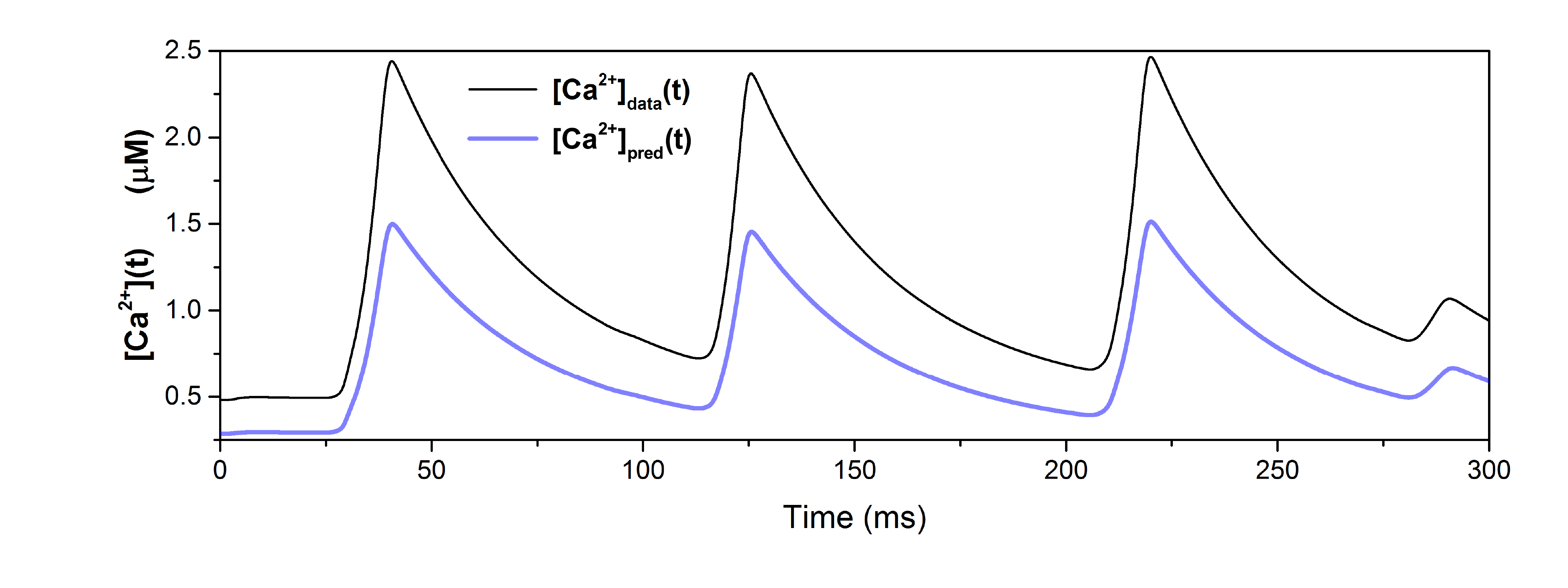} \\
\caption{Prediction for the full model with a noisy dendritic current injection (top plot). Excellent predictions arise for all state variables, of which only the voltages and calcium are shown here. Since $C_{0}$ was held at an erroneous lower value, [Ca]$(t)$ is lower in the prediction, but still exhibits bursts. There is excellent agreement in the voltage traces}
  \label{fig:soma_fixed_dendrite_dendritic_current}
\end{figure}

In fact, this method could be further extended to a systematic study in which one probes, one by one, restricted subspaces of the full model by blocking many other ion channels. The accuracy of the estimates is highly dependent on the dimension of the state and parameter space and on the associated action surface manifold, and so the data assimilation would likely be more successful in these restricted subspaces. 

On this note, we stress that the twin experiments in this study are theoretical, and the actual success of these methods using real data is still being assessed. We anticipate that difficulties will arise. For one, chemical manipulants, while highly targeted, are not perfect; K and Na channel blockers for example, can have non-neglible effects on other currents such as leak and calcium channels. The ideal situation is one in which only a single measurement of the neuron, without chemical manipulation, is needed to determine all unknown parameters and unobserved states. Without more refined data assimilation techniques, the success of this procedure is uncertain, especially as the baseline model is extended to more added currents. Nevertheless, some methods may prove promising, such as the inclusion of time-delayed measurement terms in the action, a technique we are currently studying~\cite{Rey}. 

Currently, much experimental work is being done in utilizing two-photon calcium fluorescence measurements of HVC neurons~\cite{Graber,Peh}. These measurements could prove enormously useful in state and parameter estimation, despite that they are still qualitative in nature and do not reflect great precision in actual intracellular [Ca] concentrations. From the opposite viewpoint, this method could provide a quantitatively precise correspondence between measured soma voltage and absolute [Ca]; simultaneous recordings of voltage traces and fluorescence signals coupled with the assimilation procedure outlined here would permit a precise quantification of these signals in terms of {\it absolute} calcium concentrations. This map would make fluorescence measurements particularly useful for future data assimilation procedures, expanding the set of measurable variables to also include [Ca]$(t)$.

The possibility remains that the particular stimulus used in these estimations has failed to adequately probe the phase space of the full system and that the discrepancies in the parameter estimates seen in Table \ref{tab:params_est_2} (such as $g_{Na}$) will manifest  themselves in response to other stimuli. For example, predictions may degrade in the response of the neuron to dendritic rather than somatic currents. We test this proposal by stimulating the neuron wtih a noisy dendritic current; a comparison of the system with the true parameters versus the estimated parameters is shown in Fig. \ref{fig:soma_fixed_dendrite_dendritic_current}. The traces are identical, indicating that the discrepancy in the kinetics and conductances of the Na current may be the result of either degeneracy in the model description or relative model insensivity to these parameter values, rather than an underlying failure in the data assimilation itself.

\begin{figure}
\centering
  \includegraphics[width=.5\textwidth]{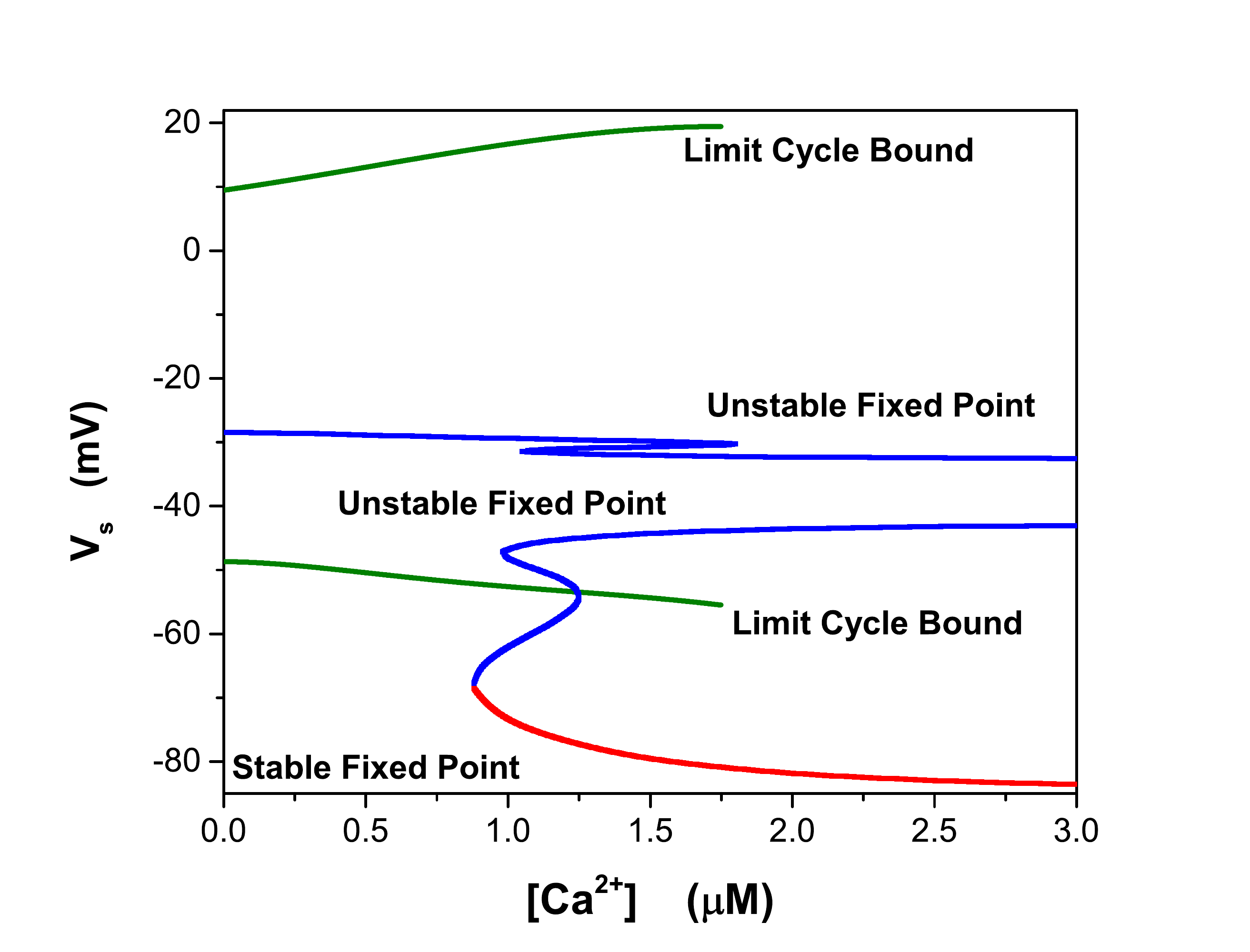} \\
  \includegraphics[width=.5\textwidth]{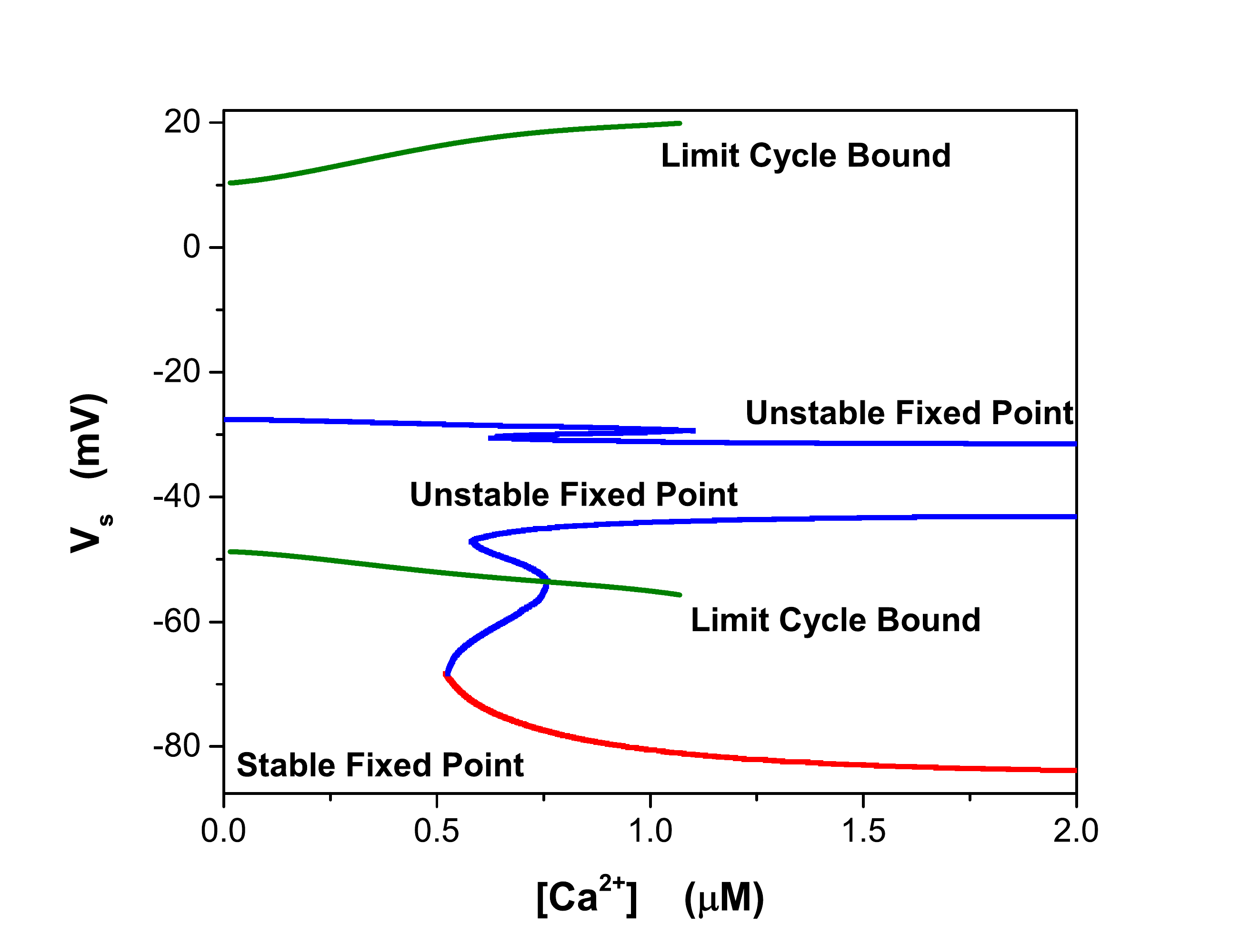}
  \caption{{\it Upper panel}. The parameters of the true trajectory, with calcium dynamics shut off, produce a bifurcation diagram in the $V_s$--[Ca] plane that exhibits regimes of stable (red) and unstable (blue) fixed points, as well as stable limit cycles, whose voltage bounds are indicated by the green lines. {\it Lower panel}. The same bifurcation diagram, again with calcium dynamics turned off, but instead using the estimated parameters rather than the true ones. The diagram is identical to the first, aside from an overall horizontal shift, indicating that identical bursting behavior can result in different parameter regimes, compensated by differences in [Ca]$(t)$}
  \label{fig:bifurcation_diagrams}
\end{figure}

We also elucidate the apparent reproducibility of the voltage traces with respect to erroneous background calcium concentrations $C\textsubscript{0}$ and  Ca\textsubscript{ext} by analyzing the bifurcation diagram of the full model without Ca dynamics. In particular, we set $\frac{d[Ca]}{dt} = 0$, and then use [Ca] as a bifurcation parameter to probe the existence and character of limit cycles and fixed points {\it as a function} of [Ca]. The bifurcation diagram of the 2D projection onto the $V_s$-[Ca] plane is shown in the upper panel of Fig. \ref{fig:bifurcation_diagrams}, using the parameters in Table \ref{tab:params_est_1}. For [Ca] between 0.8 and 1.7 $\mu$M, the system is multiply stable: stable fixed points (red) coexist with stable limit cycles (green). There is also a large region in which unstable fixed points exist (blue). 

To achieve bursting behavior, one needs calcium dynamics that modulates [Ca] periodically from below 0.8 $\mu$M, where the system only exhibits stable spiking, to above 1.7 $\mu$M, where the system exhibits only stable resting at $V_s \approx -80$ mV. This will occur for specific combinations of the parameters dictating calcium dynamics: $\phi$, $g_{Ca-L}$, $g_{Ca/K}$, $\tau_{Ca}$, $k_s$, [Ca]\textsubscript{ext}, $C\textsubscript{0}$. One set of parameters which produces such behavior corresponds to the those chosen in this paper to produce the measured data. 

Unsurprisingly, the bifurcation diagram using the {\it estimated} rather than actual parameters is nearly identical, but instead with the calcium concentration shifted (Fig. \ref{fig:bifurcation_diagrams}; lower panel). Identical  behavior can be achieved by in this case modulating the [Ca] between about 0.5 and 1.1 -- this is what is exhibited in the shifted [Ca]$(t)$ traces in Fig. \ref{fig:dendrite_only_wrong_params} and Fig. \ref{fig:soma_fixed_dendrite}. Thus, one can match the voltage traces to the model with the true parameters by compensating adjustments in the calcium dynamical parameters. This explains the discrepancies in the estimates of $g_{Ca-L}$ and $k_s$ when [Ca]\textsubscript{ext} and $C\textsubscript{0}$ are fixed to erroneous values. The model is therefore highly degenerate in the subspace of these parameters, and fixing some of these at  possibly incorrect values will cause the others to adjust accordingly to match the measured voltage traces, with little error. Importantly, it is not crucial that the guesses are precisely correct, with the understanding that the calcium concentration trace cannot be presumed to be absolute. As one is typically interested in seeing the qualitative behavior of the calcium concentrations throughout neuronal stimulation, then preserving this behavior -- along with the quantitative precision of the voltage traces -- is satisfactory.

Finally, we note that the signal-to-noise ratio of the voltage trace (about 30 dB) is chosen to reflect a typical {\it in vitro} membrane voltage recording. It is not intended as a lower limit of the acceptable SNR for the methods outlined here, and indeed we have found similarly accurate results for noisier signals, 25 dB and lower. In general, the acceptable noise for precise estimations and predictions will be a sensitive function of the model being considered, and so will vary from neuron to neuron. In addition, non-additive non-Gaussian noise, arising from effects such as synaptic input currents, would complicate this procedure. The method presented here is intended largely for {\it in vitro} measurements in which synaptic currents can be effectively set to zero. Adapting our methods for {\it in vivo} recordings is an ongoing avenue of exploration.

Similarly, the length of necessary voltage data can be kept within the limitations of typical voltage recordings, at most a few seconds. Since the injected current is chosen precisely to densely explore the model phase space, one can always tune the injected current further to allow more adequate exploration of the phase space in less time if shorter recordings are desired.

%%%%			 Figure 14

\section{Conclusion}

We have outlined a general theoretical procedure to determine a large set of parameters and state variables in RA-projecting neurons in the HVC nucleus of the avian song system, using only a somatically-injected current and a measurement of the soma membrane voltage. Neither dendritic currents, calcium concentrations, or any information of channel gating variables are needed, aside from an ansatz of their functional forms. We first proposed a baseline model of HVC\textsubscript{RA} neurons that reproduces a host of qualitative spiking features indicated by experiment, including the effects of calcium antagonists and agonists. We then used a variational approximation of path integral data assimilation, combined with a recently-described iterative annealing procedure and some straightforward experimental modification, to show that 42 unknown parameters and 6 unmeasured states in the model can be accurately estimated, with considerably lax constraints on the ranges of the unknown parameters. Importantly, this work extends previous results in that it includes the estimation of parameters that enter the model nonlinearly, such as gating time constants and threshold voltages, and it accurately predicts the time trajectory of all relevant state variables. Furthermore, we show that when this procedure is presented with actual measured data, the structure of the action level plots can readily indicate whether or not the model is incomplete. This work is an important first step in a systematic procedure to determine the fine structure of neurons, both in zebra finch HVC and elsewhere, while respecting the often stringent limitations of experiments themselves.

\bibliographystyle{spphys}       % APS-like style for physics
\bibliography{Draft_16_bib}   % name your BibTeX data base

\iffalse

\fi
\end{document}